\documentclass{article}
\usepackage{iclr2027_conference,times}

\usepackage{amsmath,amsfonts,bm}

\def\eqref#1{equation~\ref{#1}}

\def\1{\bm{1}}

\DeclareMathAlphabet{\mathsfit}{\encodingdefault}{\sfdefault}{m}{sl}
\SetMathAlphabet{\mathsfit}{bold}{\encodingdefault}{\sfdefault}{bx}{n}

\usepackage{url}
\usepackage{booktabs}
\usepackage{graphicx}
\usepackage{amsmath}
\usepackage{amssymb}
\usepackage{amsfonts}
\usepackage{multirow}
\usepackage{multicol}
\usepackage{xcolor}
\usepackage{algorithm}
\usepackage{algorithmic}
\usepackage{placeins}
\usepackage{hyperref}
\title{EVIE: Evidence-Vector-Informed Embeddings for Visual Document Retrieval}

\renewcommand{\headrulewidth}{0pt}

\author{
  Zifei Wang$^{1,*}$,\ \ Wei Wen$^{2,*\dagger}$,\ \ Qiang Ji$^{1}$,\ \ Qian-Wen Zhang$^{1,2}$,\ \ Ruizhi Qiao$^{1,2}$,\ \ Xing Sun$^{1,2}$ \\[3pt]
  $^{1}$Tencent IMA Product Center \quad $^{2}$Tencent Youtu Lab \\[3pt]
  {\normalfont\scriptsize\texttt{\{zifeiwang, jawnrwen, peasirji, cowenzhang, ruizhiqiao, winfredsun\}@tencent.com}}
}

\begin{document}

\maketitle
\vspace{-12pt}

\fancyhead{}
\lhead{}
\chead{}
\rhead{}


\begin{abstract}
Accurate and scalable visual document retrieval (VDR) requires both fine-grained page understanding and efficient indexing, yet existing approaches struggle to achieve both.
OCR-based text retrieval adds preprocessing latency and can lose visual and structural cues needed to understand complex pages.
Single-vector vision-language models bypass OCR, but compressing an entire page into one vector limits the granularity of query--document matching.
Multi-vector retrievers with MaxSim provide finer interactions, yet demand large indexes and still leave room for accuracy improvements.
We argue that overcoming these limitations requires preserving query-relevant page evidence throughout representation learning and index construction.
To this end, we introduce \textbf{\textit{EVIE}} (Evidence-Vector-Informed Embeddings), a family of native visual document retrievers integrating three key innovations:
(1) Evidence-judged data governance, which uses a multimodal judge to identify answer-bearing positives and filter unreliable negatives.
(2) Bidirectional teacher--student learning with symmetric listwise distillation and prefix-based Matryoshka representation learning (Prefix-MRL), enabling one student checkpoint to serve six nested embedding dimensions without re-encoding.
(3) Hierarchical agglomerative index compression (HAC), which clusters page tokens with spatial regularization and stores semantic centroids for single-stage MaxSim retrieval.
Extensive experiments across 138 tasks from ViDoRe V1, V2, V3, and JinaVDR validate EVIE.
EVIE-8B achieves 66.75 nDCG@10 on V3, exceeding the best external baseline by 1.43 points, with a four-suite average of 79.51.
EVIE-4.5B with HAC retains 59.58 nDCG@10 at only 3.81 GiB per million pages, reducing vector payload by $128\times$ relative to an uncompressed 1024-token, 256-dimensional BF16 index.
Together, these results improve the accuracy--storage trade-off for visual document retrieval.
\end{abstract}

\vspace{-8pt}
\begingroup
\setlength{\intextsep}{4pt}
\begin{figure}[H]
\centering
\includegraphics[width=\linewidth]{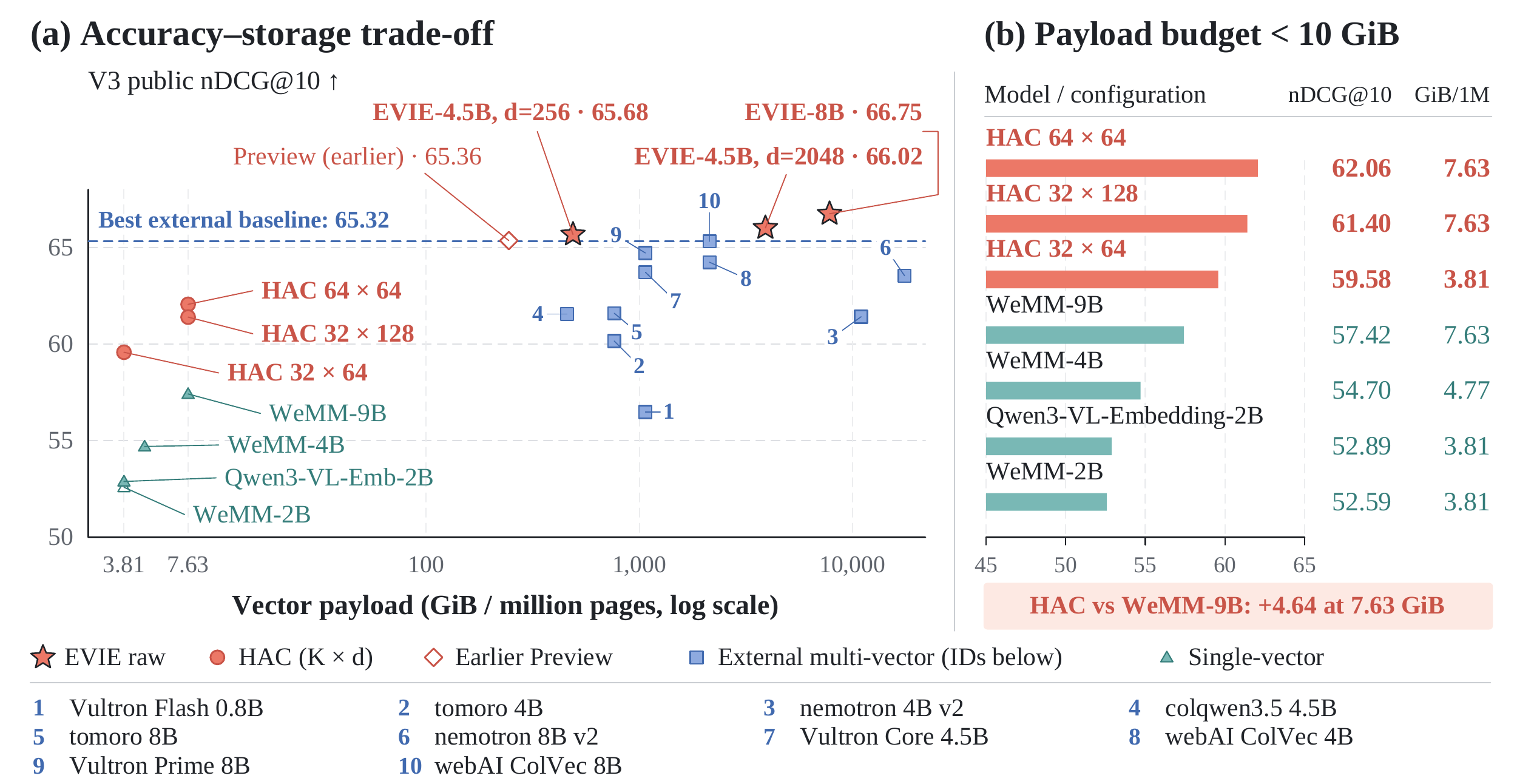}
\small
\caption{\textbf{Accuracy versus vector payload on public ViDoRe V3.} (a) All configurations in Table~\ref{tab:vidore_v3}; (b) Payload $<10$ GiB/1M pages; bars start at 45. HAC: $K\times d$.}
\label{fig:pareto_curve}
\end{figure}
\endgroup

\section{Introduction} \label{sec:intro}

\begin{figure}[htbp]
\centering
\includegraphics[width=\linewidth]{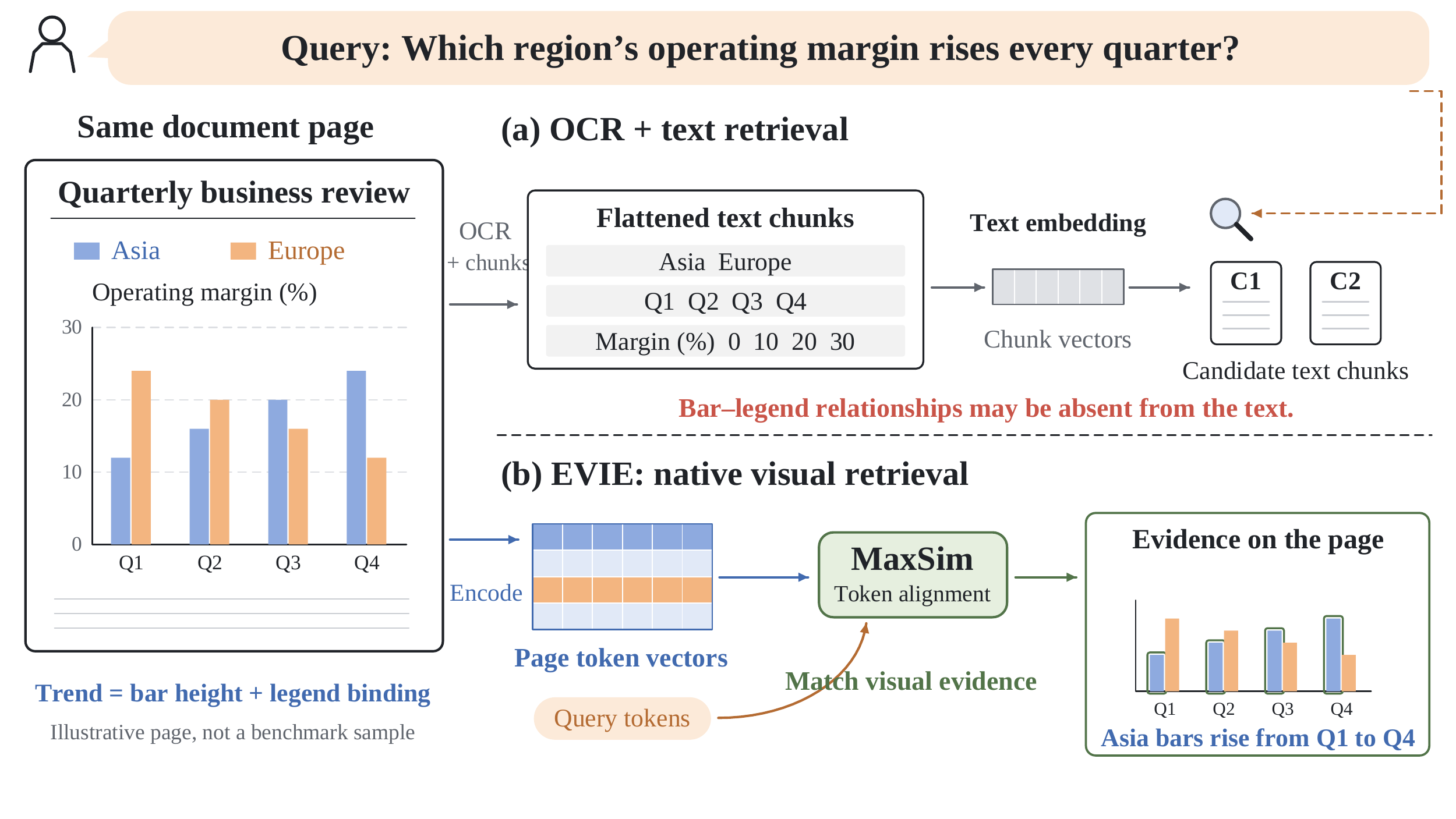}
\caption{\textbf{Motivation: retrieving graphical evidence beyond OCR text.} The same query and constructed chart are routed through (a) OCR, text chunking, and text retrieval, or (b) native visual encoding and token-level MaxSim. The example shows how flattened text may preserve words and axis ticks while omitting bar heights and colour-to-series binding; the page image retains the rising Asia series as visual evidence.}
\label{fig:ocr_motivation}
\end{figure}

Retrieving evidence from visually rich documents is fundamental to document search and Retrieval-Augmented Generation (RAG). Financial reports, scientific articles, and scanned records convey information through both text and the arrangement of tables, figures, and page elements. Conventional OCR-based retrieval converts pages into text through recognition, layout parsing, and chunking, adding preprocessing overhead while potentially losing visual relationships \citep{colpali2024}. For example, identifying a region whose operating margin rises each quarter requires associating bar heights with time and the legend; region names and axis labels alone are insufficient (Figure~\ref{fig:ocr_motivation}).

Native vision--language encoders instead represent page images directly. Single-vector models, such as DSE and Qwen3-VL-Embedding \citep{dse2024,qwen3vl_embed2025}, map queries and pages into a shared embedding space, avoiding OCR and yielding compact indexes. However, summarizing many distinct page regions in one vector can obscure local differences between otherwise similar pages. Multi-vector late-interaction models, exemplified by ColPali \citep{colpali2024}, retain separate patch representations and match query tokens to them through MaxSim \citep{colbert2020}, enabling finer alignment with visual evidence.

Fine-grained matching nevertheless leaves a coupled learning and storage problem. Visually similar pages may provide different answers, while mined hard negatives may contain complete or partial answer evidence; treating them uniformly as negatives can suppress useful matches. At EVIE's configured maximum of 1024 image tokens per page, a million pages of 256-dimensional BF16 image vectors require approximately 488 GiB of vector payload. Narrowing vectors and merging tokens reduce this cost, but both can alter the local similarities and best-matching regions used by MaxSim. A model trained at one width therefore need not preserve its rankings at narrower widths, and token compression must retain local distinctions across queries. Using a compact index only for candidate retrieval does not remove the storage burden when reranking still relies on stored full token representations, and missed candidates remain unrecoverable. Accurate, compact retrieval thus requires reliable evidence supervision, representations trained across widths, and compressed tokens that directly support final relevance scoring.

To address these challenges, we introduce \textbf{EVIE} (\textbf{\underline{E}}vidence-\textbf{\underline{V}}ector-\textbf{\underline{I}}nformed \textbf{\underline{E}}mbeddings), a family of native visual document retrievers combining evidence-judged supervision with compact multi-vector indexing. Its \emph{evidence-judged data governance} uses a multimodal judge to promote candidates with complete answer evidence to positives and exclude those with partial evidence from the negative loss. In our audit, these two categories account for 14.4\% of mined candidate negative slots according to the judge. Using this supervision, an unmasked bidirectional teacher, \emph{EVIE-8B}, transfers its rankings to \emph{EVIE-4.5B} through symmetric bidirectional listwise distillation. The student incorporates \emph{Prefix-MRL}, adapting Matryoshka Representation Learning \citep{kusupati2022matryoshka} to late interaction by applying retrieval supervision and ranking distillation at every prefix width. One projection head consequently serves six nested dimensions from 64 to 2048 without re-encoding.

\begin{figure}[htbp]
\centering
\includegraphics[width=\linewidth]{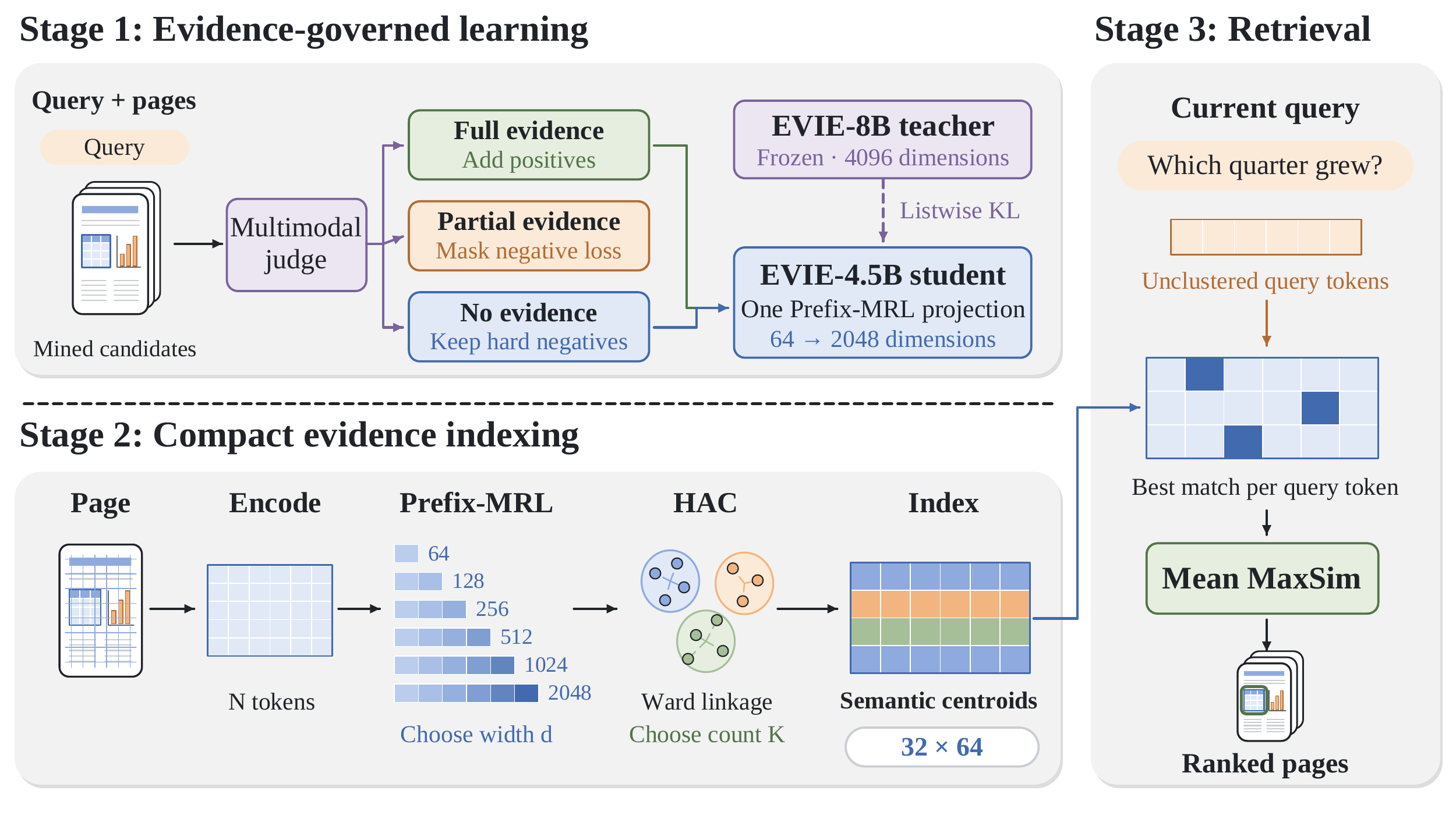}
\caption{\textbf{EVIE training, indexing, and retrieval.} Evidence-judged supervision trains the Prefix-MRL student; HAC compresses page tokens into semantic centroids for direct MaxSim scoring. Page contents and representations are schematic.}
\label{fig:evie_overview}
\end{figure}

\emph{HAC} reduces token counts through offline hierarchical agglomerative clustering. Position-only pooling may merge heterogeneous content, while purely semantic grouping may conflate similar elements in distant regions. HAC therefore uses spatial information to guide assignments but computes stored centroids solely from the original semantic vectors. These centroids support direct MaxSim scoring without an uncompressed index or reranking pass (Figure~\ref{fig:evie_overview}).

Experiments across 138 tasks from ViDoRe V1, V2, V3, and JinaVDR evaluate EVIE's accuracy--storage trade-off (Figure~\ref{fig:pareto_curve}). EVIE-8B reaches 66.75 nDCG@10 on the eight public ViDoRe V3 domains, exceeding the strongest compared external baseline by 1.43 points. EVIE-4.5B reaches 66.02 at $d=2048$, within 0.73 points of the teacher. At the same 7.63 GiB vector payload per million pages, EVIE-4.5B with HAC scores 62.06 versus 57.42 for WeMM-9B under the same evaluation protocol, a gain of 4.64 points. Further compression retains 59.58 at 3.81 GiB, a $128\times$ vector-payload reduction relative to EVIE's configured maximum of 1024 image vectors per page at 256 dimensions and two bytes per component.

The main contributions of this paper are as follows:
\begin{itemize}
\item We introduce evidence-judged supervision that grades mined pages by answer completeness, expands positive supervision, and excludes partial-evidence candidates from the negative loss.
\item We develop a bidirectional teacher--student framework with Prefix-MRL, training six nested widths with retrieval and ranking objectives in one checkpoint without re-encoding.
\item We propose HAC, separating spatially regularized cluster assignment from semantic centroid storage to support single-stage MaxSim retrieval over compressed page tokens.
\end{itemize}

\section{Related Work}

EVIE builds on visual document representation learning, retrieval supervision, and compact multi-vector indexing.

\subsection{Visual Document Retrieval}
Native visual document retrievers encode page images directly to preserve information that may be lost during text extraction. DSE and Qwen3-VL-Embedding \citep{dse2024,qwen3vl_embed2025} produce a single vector per page, offering compact indexes. ColPali \citep{colpali2024} applies ColBERT-style late interaction \citep{colbert2020} to patch representations, enabling query-token matching to individual page regions. Jina-embeddings-v4 \citep{jina_v4_2025} supports both single-vector and multi-vector outputs. NeMoRetriever-ColEmbed and Nemotron-ColEmbed-V2 \citep{nemoretriever2024,nemotron_colembed_v2} extend multimodal late interaction with larger vision--language backbones, while ModernVBERT \citep{modernvbert2025} studies efficient bidirectional document encoding. MoCa \citep{moca2025} investigates modality-aware continual pretraining for bidirectional multimodal embeddings. Universal encoders such as E5-V, GME, and UniME \citep{e5v2024,gme2024,unime2024} address broader multimodal retrieval, while the VLM2Vec models and associated MMEB benchmarks \citep{vlm2vec2024,vlm2vec_v2_2025} cover diverse tasks involving images, videos, and visual documents. EVIE follows the bidirectional late-interaction architecture and focuses on the accuracy--storage trade-off in page retrieval.

\subsection{Supervision and Distillation for Retrieval}
Hard-negative mining strengthens contrastive retrieval training, but retrieved candidates may include unlabeled positives or partially relevant documents. ColBERTv2 \citep{colbertv2_2022} combines mined negatives with cross-encoder supervision and KL distillation to improve training signals, while Unveil \citep{unveil2025} distills visual--textual knowledge into a visual document retriever. ColMate \citep{colmate2025} enriches contrastive training with masked text supervision linked to document regions. mmE5 \citep{mme5_2024} emphasizes high-quality synthetic training data, while Embed-RL and TRACE \citep{embedrl2025,trace2025} explore reasoning-driven multimodal embedding learning. EVIE makes answer completeness an explicit criterion for candidate relabeling: a multimodal judge promotes full-evidence candidates to positives, excludes partial-evidence candidates from the negative loss, and retains no-evidence candidates as negatives. This supervision trains a bidirectional teacher whose query-to-document and document-to-query ranking distributions are subsequently transferred to the student. Evidence grading determines how mined candidates contribute to retrieval training, while teacher distillation supplies ranking supervision across embedding widths.

\subsection{Variable-Dimension Multi-Vector Representations}
Matryoshka Representation Learning (MRL) \citep{kusupati2022matryoshka} trains nested embedding prefixes so that one representation supports multiple output dimensions. Qwen3-VL-Embedding \citep{qwen3vl_embed2025} applies this principle to multimodal embeddings, and ColNetraEmbed \citep{m3dr2025} extends dimension flexibility to multi-vector document retrieval. In late interaction, truncation changes the token similarities used by MaxSim, making retrieval quality at each width a training concern. EVIE's Prefix-MRL therefore applies both retrieval supervision and bidirectional listwise distillation at every trained prefix width. A shared projection head serves six nested dimensions from 64 to 2048 without rerunning the backbone, allowing index width to be selected from one student checkpoint.

\subsection{Multi-Vector Index Compression}
Multi-vector index size can be reduced by compressing individual vectors or storing fewer vectors per document. ColBERTv2 \citep{colbertv2_2022} uses residual compression to reduce storage per vector, and ColPali \citep{colpali2024} evaluates hierarchical mean token pooling to merge redundant patch representations. PLAID \citep{plaid2022} accelerates late-interaction retrieval through centroid interaction and pruning, illustrating that serving efficiency also depends on the search engine. ColParse \citep{colparse2025} constructs compact representations from layout-informed regions; CausalEmbed \citep{causalembed2025} generates autoregressive latent tokens; and MetaEmbed \citep{metaembed2025} varies the number of learned meta-tokens. HAC follows the token-pooling approach and uses 2D spatial regularization in hierarchical clustering to account for page layout when grouping semantically similar patches. Cluster assignments use joint semantic--spatial features, whereas stored centroids average only the original semantic vectors. These centroids directly support MaxSim scoring. Combined with Prefix-MRL, this permits independent choices of embedding width and token count without retaining an uncompressed index for reranking.

\section{Model Architecture and Late Interaction Representation} \label{sec:representation}

EVIE encodes queries and page images independently into token vectors and scores relevance through late interaction. This matching interface connects page-level supervision, retrieval at different embedding widths, and the compressed index developed in Sections~\ref{sec:training}--\ref{sec:hac}.

\subsection{Input Sequences and Contextual Encoding}
Pages undergo aspect-preserving dynamic resizing, $16\times16$ patch embedding, and $2\times2$ spatial merging, yielding $N\le1024$ image tokens. The processor records the resulting two-dimensional grid and its raster ordering; HAC derives page coordinates from this metadata without OCR or layout parsing. The page sequence also contains the fixed image-description prompt and special tokens, which participate in raw retrieval but are excluded from HAC pooling. Queries use the checkpoint's BPE tokenizer with ten appended \texttt{<|endoftext|>} augmentation tokens. These are valid tokens despite sharing the padding token ID, so validity is determined by the processor's attention mask.

Within each model, query and page encoders share parameters. For input $x\in\{q,p\}$, let $H_x\in\mathbb R^{L_x\times h}$ denote the hidden-state matrix and $a_x\in\{0,1\}^{L_x}$ its valid-token mask; $L_x$ includes batch padding, and $L_p$ includes both image and context tokens. To expose tokens to context on both sides, we remove causal masking from the eight full-attention layers of each 32-layer language backbone \citep{nemotron_colembed_v2}. The other 24 linear/recurrent layers remain unchanged. In the full-attention layers, the additive mask changes as
\begin{equation}
A_{\mathrm{causal}}=A_{\mathrm{pad}}+A_{\mathrm{tri}}
\quad\longrightarrow\quad A_{\mathrm{EVIE}}=A_{\mathrm{pad}},
\label{eq:attn}
\end{equation}
where $A_{\mathrm{tri}}$ blocks future positions and $A_{\mathrm{pad}}$ blocks padding. Bidirectional context is confined to each independently encoded input; query--page interaction occurs only during scoring.

\subsection{Model Initialization and Prefix Projection}
EVIE-8B initializes from \texttt{Qwen/Qwen3.5-9B}\footnote{\nolinkurl{https://huggingface.co/Qwen/Qwen3.5-9B}}. EVIE-4.5B initializes from the retrieval checkpoint \texttt{tencent/EVIE-Preview-4.5B}\footnote{\nolinkurl{https://huggingface.co/tencent/EVIE-Preview-4.5B}}, whose backbone is \texttt{Qwen/Qwen3.5-4B}\footnote{\nolinkurl{https://huggingface.co/Qwen/Qwen3.5-4B}}. Both backbones belong to the Qwen3.5 family \citep{qwen35_2026}.
Their hidden and maximum retrieval widths $(h,D_{\max})$ are $(4096,4096)$ and $(2560,2048)$, respectively. The deployed models contain 8.41B and 4.54B parameters, including the retrieval head after LoRA merging. Both train rank-32 LoRA adapters \citep{hu2021lora} with $\alpha=128$ on language attention, recurrent, and feed-forward projections, together with the full retrieval head; the visual encoder remains frozen.

\begin{figure}[htbp]
\centering
\includegraphics[width=\linewidth]{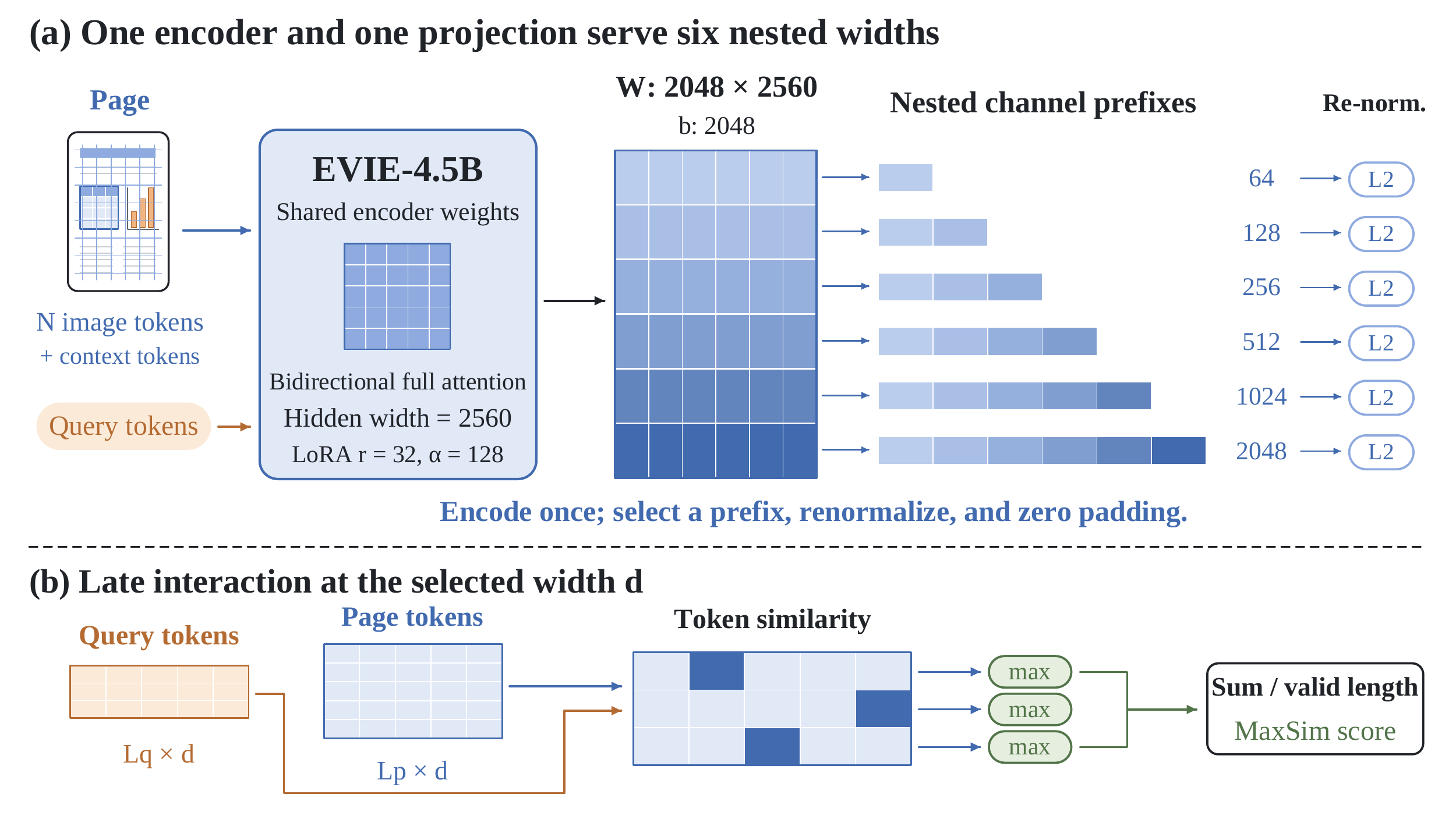}
\caption{\textbf{EVIE-4.5B encoding and prefix-wise matching.} Independently encoded inputs share an affine projection. Each selected prefix is normalized and padding rows are zeroed before MaxSim. Page sequences include image and context tokens; $L_q$ and $L_p$ denote padded sequence lengths. Prefix blocks and matrices are schematic.}
\label{fig:architecture}
\end{figure}

Each model uses a shared affine projection $W\in\mathbb R^{D_{\max}\times h}$ with bias $b\in\mathbb R^{D_{\max}}$ for query and page tokens:
\begin{equation}
\begin{aligned}
U_x &= H_xW^\top+\mathbf 1b^\top\in\mathbb R^{L_x\times D_{\max}},\\
E_x^{(d)} &= \operatorname{diag}(a_x)\,\mathrm{RowL2}(U_x[:,:d])\in\mathbb R^{L_x\times d}.
\end{aligned}
\label{eq:prefix}
\end{equation}
Here $\mathrm{RowL2}$ normalizes each token vector, with zero vectors remaining zero; padding rows are then zeroed by $a_x$. The teacher uses $d=4096$. The student supports $d\in\mathcal D=\{64,128,256,512,1024,2048\}$ through nested prefixes, retaining the Preview head's first 128 weight rows and bias entries at initialization. Available full-width projections can supply every prefix through slicing and renormalization without rerunning the backbone (Figure~\ref{fig:architecture}).

\subsection{Late Interaction Scoring (MaxSim)}
Writing $e_{x,i}^{(d)}$ for row $i$ of $E_x^{(d)}$, the query-normalized score at width $d$ is
\begin{equation}
s_d(q,p)=\frac{1}{n_q}\sum_{i:a_{q,i}=1}\max_{j:a_{p,j}=1}
\left\langle e_{q,i}^{(d)},e_{p,j}^{(d)}\right\rangle,
\qquad n_q=\sum_{i=1}^{L_q}a_{q,i}.
\label{eq:maxsim}
\end{equation}
\begin{samepage}
Before taking the maximum, invalid page positions are masked with a dtype-safe negative value. Query padding positions are excluded from the aggregation. Student training uses the query mean above; evaluation uses the sum, which gives the same ranking for a fixed query.
Nested projection parameters alone do not ensure useful rankings at every width. Section~\ref{sec:training} specifies the relevance labels and teacher signals that train this matching function at each student prefix; Section~\ref{sec:hac} then replaces page tokens with centroids for compact retrieval.
\par
\end{samepage}

\section{Training and Distillation Framework} \label{sec:training}

EVIE learns evidence-sensitive MaxSim rankings in two stages: retrieval supervision establishes a teacher, and its ranking distributions and positive--negative margins supervise every student prefix. An additional constraint preserves the student's initial 128-dimensional retrieval geometry.

\subsection{Evidence-Judged Data Governance} \label{sec:governance}
To avoid penalizing answer-bearing pages, a multimodal judge, GLM-5V-Turbo \citep{glm5v_turbo_2026}, grades mined candidates by answer completeness: \textbf{full-evidence} pages augment the registered positives, \textbf{partial-evidence} pages are excluded from their negative slots, and \textbf{no-evidence} pages remain negatives. The corpus contains 775,635 original query groups, including 179,295 records without evidence judgments; unjudged candidates retain their original negative status. Before positive expansion, 223,877 of 1,551,270 candidate slots (14.4\%) are masked, and 175,137 query groups have multiple positives.

Let $\mathcal P_g$ be the positive-page set of group $g$. The all-positive training view expands each group into one record $i$ per positive $p_i\in\mathcal P_{g(i)}$, yielding 1,050,270 records with weights $\omega_i=1/|\mathcal P_{g(i)}|$. Each group's nominal weights therefore sum to one. Each record also carries two negative slots $n_{ik}$ and binary validity masks $m_{ik}$; full- and partial-evidence slots have $m_{ik}=0$. The expanded view contains 462,897 masked slots out of 2,100,540. Group and page identities accompany these records to prevent known positives from becoming competing negatives.
These page-level judgments supervise relevance without providing patch-level localization labels. The counts describe the judge's decisions; Section~\ref{sec:controlled_validation} evaluates their agreement with human labels and the retrieval gains from each governance step.

\subsection{Retrieval Supervision and Teacher Construction}
At an active width $d$, abbreviate $s_d$ as $s$ and write $s_{ij}=s(q_i,p_j)$, $s_i^+=s(q_i,p_i)$, and $\Delta_{ik}=s_i^+-s(q_i,n_{ik})$. Retrieval supervision combines masked negative ranking with in-batch cross-entropy:
\begin{equation}
\begin{aligned}
\ell_{\mathrm{hard},i}&=\frac{\sum_k m_{ik}\,\mathrm{softplus}(-\Delta_{ik}/\tau)}{\max(1,\sum_k m_{ik})},\\
\ell_{\mathrm{CE},i}&=-\log\frac{\exp(s_i^+/\tau)}{\sum_{j\in\mathcal C_i}\exp(\widetilde s_{ij}/\tau)},\qquad
\ell_{\mathrm{ret},i}=\tfrac12(\ell_{\mathrm{hard},i}+\ell_{\mathrm{CE},i}).
\end{aligned}
\label{eq:ret}
\end{equation}
Here $\tau=0.02$, and $\mathcal C_i$ contains permitted entries from the gathered positive-page batch, including $p_i$. To attenuate possible false negatives, $\widetilde s_{ij}=0.5s_{ij}$ for non-target entries satisfying $s_{ij}>0.95s_i^+$; other scores are unchanged. Empty valid-negative sets contribute zero ranking loss. For weighted records, each rank reduces losses as $\mathcal R_\omega(\ell)=\sum_{i\in\mathcal B_r}\omega_i\ell_i/\sum_{i\in\mathcal B_r}\omega_i$, followed by DDP gradient averaging.

Two 4096-dimensional teacher branches share initialization and retrieval supervision. The sampled-positive branch draws one page uniformly from $\mathcal P_g$ per record, averaging ranking loss over all valid slots on each rank and CE over records. The all-positive branch uses $\mathcal R_\omega(\ell_{\mathrm{ret}})$ and excludes other records of the same query group from $\mathcal C_i$. After training, their full model parameters are combined as $\theta_T=0.4\theta_{\mathrm{sampled}}+0.6\theta_{\mathrm{all}}$ and frozen for student supervision.

\subsection{Prefix-wise Retrieval and Distillation}
The student uses the all-positive view. At each width $d\in\mathcal D$, retrieval labels establish relevance, teacher distributions transfer candidate rankings, and teacher margins constrain difficult positive--negative pairs. The complete objective is
\begin{equation}
\mathcal L_S=\overline{\mathcal L}_{\mathrm{ret}}+
a_t\left(\overline{\mathcal L}_{\mathrm{rel}}+0.25\overline{\mathcal L}_{\mathrm{mar}}+0.25\mathcal L_{\mathrm{anc}}^{(128)}\right),
\label{eq:student}
\end{equation}
where $\overline{\mathcal L}_x=|\mathcal D|^{-1}\sum_{d\in\mathcal D}\mathcal R_\omega(\ell_x^{(d)})$ gives equal weight to all six prefixes, and $a_t$ calibrates the combined distillation strength.

\paragraph{Candidate distributions and masks.}
Query-to-page distillation uses gathered positive pages and explicit negative slots from all ranks. Page-to-query distillation normalizes the same positive-page score matrix over gathered queries. In that matrix, non-target entries sharing a query group or matching a positive in $\mathcal P_{g(i)}$ are masked for student CE and both KL directions. The expanded query-to-page pool additionally masks known positives and invalid negative slots. Teacher and student use identical masks; partial-evidence exclusions are limited to annotated slots. Softmax of remaining MaxSim scores at $\tau_{\mathrm{kd}}=0.13$ defines $P_T^\rho$ and $P_S^{(d),\rho}$ for $\rho\in\{q\!\to\!p,p\!\to\!q\}$, conditioned on $q_i$ or $p_i$, respectively.

\paragraph{Ranking transfer and anchor preservation.}
The two directions receive equal weight:
\begin{equation}
\ell_{\mathrm{rel},i}^{(d)}=\tfrac12\sum_\rho c_i^\rho\,
\mathrm{KL}\!\left(P_{T,i}^\rho\parallel P_{S,i}^{(d),\rho}\right).
\label{eq:kd}
\end{equation}
Symmetry refers to these retrieval directions: both KL terms run from teacher to student and use the same asymmetric MaxSim score function. The scale $a_t$ controls distillation strength independently of temperature. For $V$ valid candidates, confidence is $c=0.1+0.9[1-H(P_T)/\log\max(2,V)]$, where $H$ denotes entropy; $c$ is multiplied by $0.25$ when the teacher's top-ranked entry is not the designated target. This attenuates uncertain or target-inconsistent supervision. Explicit negatives additionally transfer score margins through
\begin{equation}
\ell_{\mathrm{mar},i}^{(d)}=
\frac{\sum_k m_{ik}\,\sigma(\Delta_{T,ik}/\tau_{\mathrm{kd}})\,\mathrm{smoothL1}(\Delta_{S,ik}^{(d)},\Delta_{T,ik})}{\max(1,\sum_k m_{ik})},
\label{eq:margin}
\end{equation}
where $\sigma$ is the sigmoid function and Smooth-L1 uses unit transition scale. To retain the initial d128 geometry, $\mathcal L_{\mathrm{anc}}^{(128)}$ applies the weighted bidirectional KL of Eq.~(\ref{eq:kd}) with $c=1$, using the frozen EVIE-Preview-4.5B as target and only in-batch positive pages. This anchor is evaluated in a separate no-gradient pass with the student's adapters disabled.

\begin{algorithm}[htbp]
\caption{EVIE-4.5B Prefix-wise Training Step}
\label{alg:training}
\footnotesize
\begin{algorithmic}[1]
\REQUIRE Records $(q_i,p_i,\{n_{ik},m_{ik}\}_k,\omega_i,g(i),\mathcal P_{g(i)})$; frozen teacher $T$; student $S$; widths $\mathcal D$
\STATE Gather group/page identities; construct CE and row/column KL masks
\STATE Encode queries, positives, and negative slots once through the student backbone: $H_q,H_p,H_n$
\STATE Without gradients, score the same inputs with $T$ and query/positive inputs with the adapter-disabled d128 anchor
\FORALL{$d \in \mathcal{D}$}
    \STATE Project each token matrix $H$: $E^{(d)}\gets\mathrm{L2}_{\mathrm{row}}(H W_{:d}^{\top})$; compute gathered MaxSim scores
    \STATE Apply masks and compute $\mathcal R_\omega(\ell_{\mathrm{ret}}^{(d)}),\mathcal R_\omega(\ell_{\mathrm{rel}}^{(d)}),\mathcal R_\omega(\ell_{\mathrm{mar}}^{(d)})$
\ENDFOR
\STATE Average each loss across widths; compute $\mathcal L_{\mathrm{anc}}^{(128)}$ with the same weighted reduction
\STATE Calibrate $a_t$ during warmup; combine losses by Eq.~(\ref{eq:student})
\STATE Backward and update student parameters, respecting the projection warmup schedule
\end{algorithmic}
\end{algorithm}

\begin{figure}[htbp]
\centering
\includegraphics[width=\linewidth]{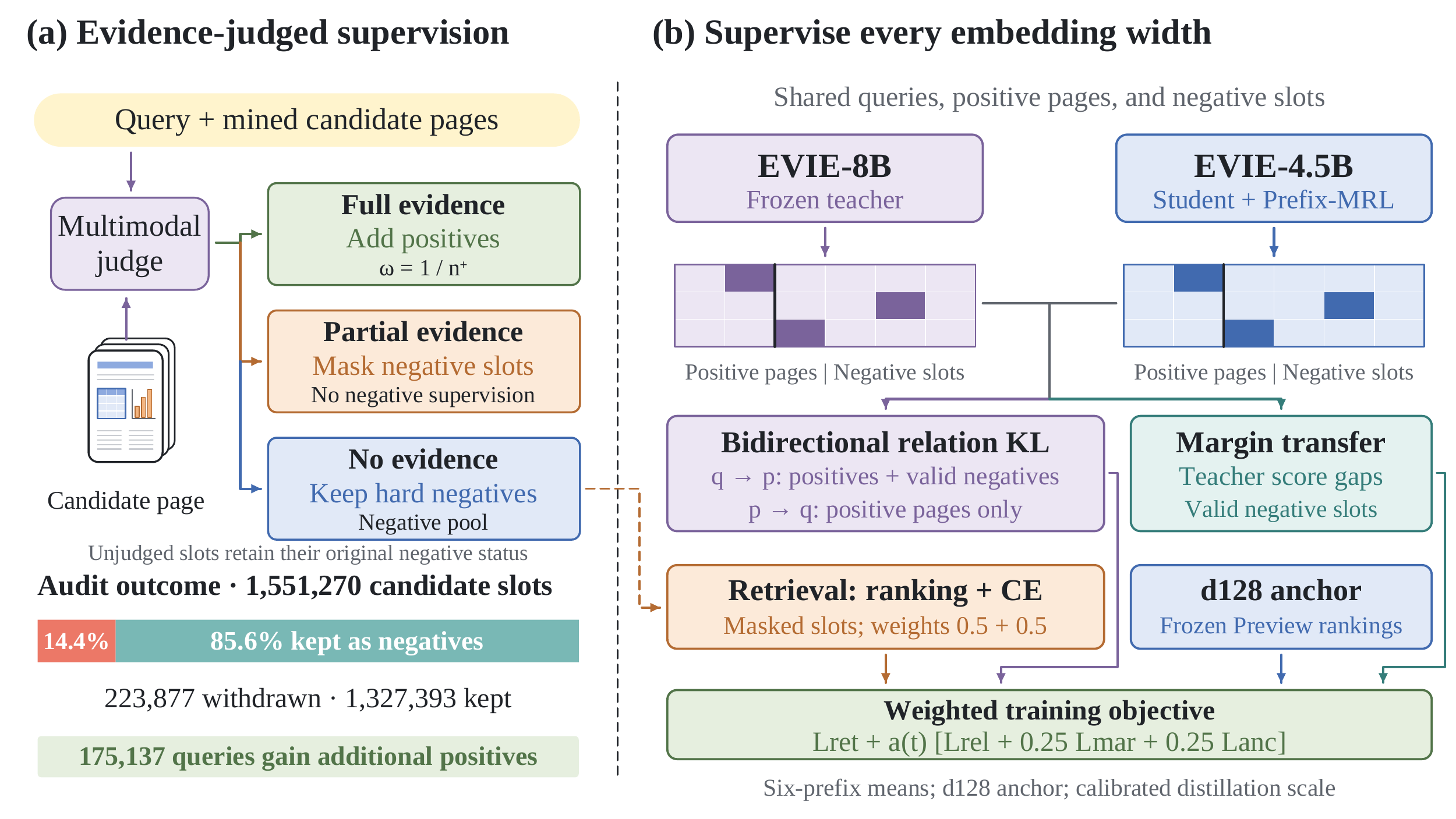}
\caption{\textbf{Evidence supervision and prefix-wise ranking transfer.} Judgments define positive records and negative-slot masks. At every width, the student learns retrieval labels, teacher ranking distributions, and positive--negative margins; a frozen Preview anchor constrains d128. Explicit negatives extend query-to-page distillation, while page-to-query distributions use positive pages. The audit precedes record expansion; matrices are schematic.}
\label{fig:training_distillation}
\end{figure}

\subsection{Optimization Details}
The student initializes from EVIE-Preview-4.5B, retaining its first 128 projection weight rows and bias entries and initializing the additional channels. Only added channels update for the first 100 optimizer steps; language-backbone LoRA and the full projection then train jointly. During these 100 steps, $a_t$ is calibrated every ten steps using query/positive d128 representation gradients: the target is $0.5\|\nabla\overline{\mathcal L}_{\mathrm{ret}}\|/\|\nabla\mathcal L_{\mathrm{aux}}\|$, clipped to $[0.05,20]$, where $\mathcal L_{\mathrm{aux}}$ is the bracketed term in Eq.~(\ref{eq:student}). Gradient norms aggregate squared magnitudes across ranks; the first target initializes $a_t$, subsequent targets use EMA decay $0.9$, and the resulting scale is held fixed thereafter.

Each teacher branch and the student train for one epoch on 32 GPUs with two records per GPU and eight accumulation steps: the gathered contrastive batch is 64, and the effective optimization batch is 512. Each student micro-batch is drawn from one data source. AdamW uses weight decay $0.02$, cosine decay with 8\% learning-rate warmup, and peak rates $4.57\times10^{-5}$ for the teacher and $1.5\times10^{-5}$ for the student. The visual token budget is 1024; LoRA uses rank 32, $\alpha=128$, and dropout 0.197. Figure~\ref{fig:training_distillation} summarizes training on uncompressed tokens at each prefix. HAC subsequently reduces the stored page vectors while retaining the query representation and MaxSim aggregation.

\section{HAC: Hierarchical Agglomerative Index Compression} \label{sec:hac}

After training, HAC performs query-agnostic offline compression at a selected Prefix-MRL width $d$, reducing the number of stored page vectors while retaining direct MaxSim scoring. It forms clusters using semantic and positional features, then stores normalized means of the original retrieval vectors.

\subsection{Position-Guided Clustering and Semantic Centroids}
HAC extends hierarchical mean token pooling in ColPali \citep{colpali2024} with page-grid positional features for cluster assignment at the selected prefix width. Whereas ColParse \citep{colparse2025} encodes parsed sub-images and combines their vectors with a global-page representation, HAC pools existing page tokens using the processor grid, without region parsing or re-encoding.

HAC takes the image-token rows of $E_p^{(d)}$ as $v_1,\ldots,v_N\in\mathbb R^d$ in raster order, for $d\in\mathcal D$; raw evaluation additionally retains non-image context tokens. For a single-image grid $(1,g_h,g_w)$, the visual encoder's $2\times2$ merge yields $H=g_h/2$, $W=g_w/2$, and $N=HW$ image tokens. In raster order, token $i$ at row $r$ and column $c$ has coordinates $x_i=c/\max(W-1,1)$ and $y_i=r/\max(H-1,1)$, with zero-based indices.

To condition assignments on page position, we use $m=d/4$ frequencies per coordinate, $\gamma_k=10000^{k/(m-1)}$ for $k=0,\ldots,m-1$. The four sinusoidal blocks are concatenated and normalized:
\begin{equation}
\begin{aligned}
\widetilde p_i &= \left[(\sin(\gamma_kx_i))_k;\ (\cos(\gamma_kx_i))_k;\ (\sin(\gamma_ky_i))_k;\ (\cos(\gamma_ky_i))_k\right]^\top,\\
p_i &= \mathrm{L2}(\widetilde p_i)\in\mathbb R^d.
\end{aligned}
\label{eq:pos}
\end{equation}
The clustering feature combines unit-norm semantic and positional vectors:
\begin{equation}
z_i = \mathrm{L2}\left( (1 - w_{\text{pos}}) \mathrm{L2}(v_i) + w_{\text{pos}} p_i \right),
\label{eq:zfeat}
\end{equation}
We fix $w_{\text{pos}}=0.1$ (Appendix~\ref{app:extension_controls}).
Section~\ref{sec:controlled_validation} compares this setting with semantic-only clustering on the final student. Euclidean Ward linkage \citep{ward1963} repeatedly merges the cluster pair with the smallest increase in within-cluster squared error over $z_i$, stopping at $K'=\min(K,N)$ clusters.
Cluster membership determines which original retrieval vectors enter each centroid:
\begin{equation}
\mu_c = \mathrm{L2}\left( \frac{1}{|c|} \sum_{i \in c} v_i \right).
\label{eq:centroid}
\end{equation}
No additional positional feature is averaged into $\mu_c$. Queries use the same prefix width, remain unclustered, and score directly against these centroids by MaxSim. Thus the query representation and scoring rule are unchanged, although pooling can alter the best-matching vector and its score. The raw-to-HAC comparison also includes removal of the context tokens.

\begin{figure}[htbp]
\centering
\includegraphics[width=\linewidth]{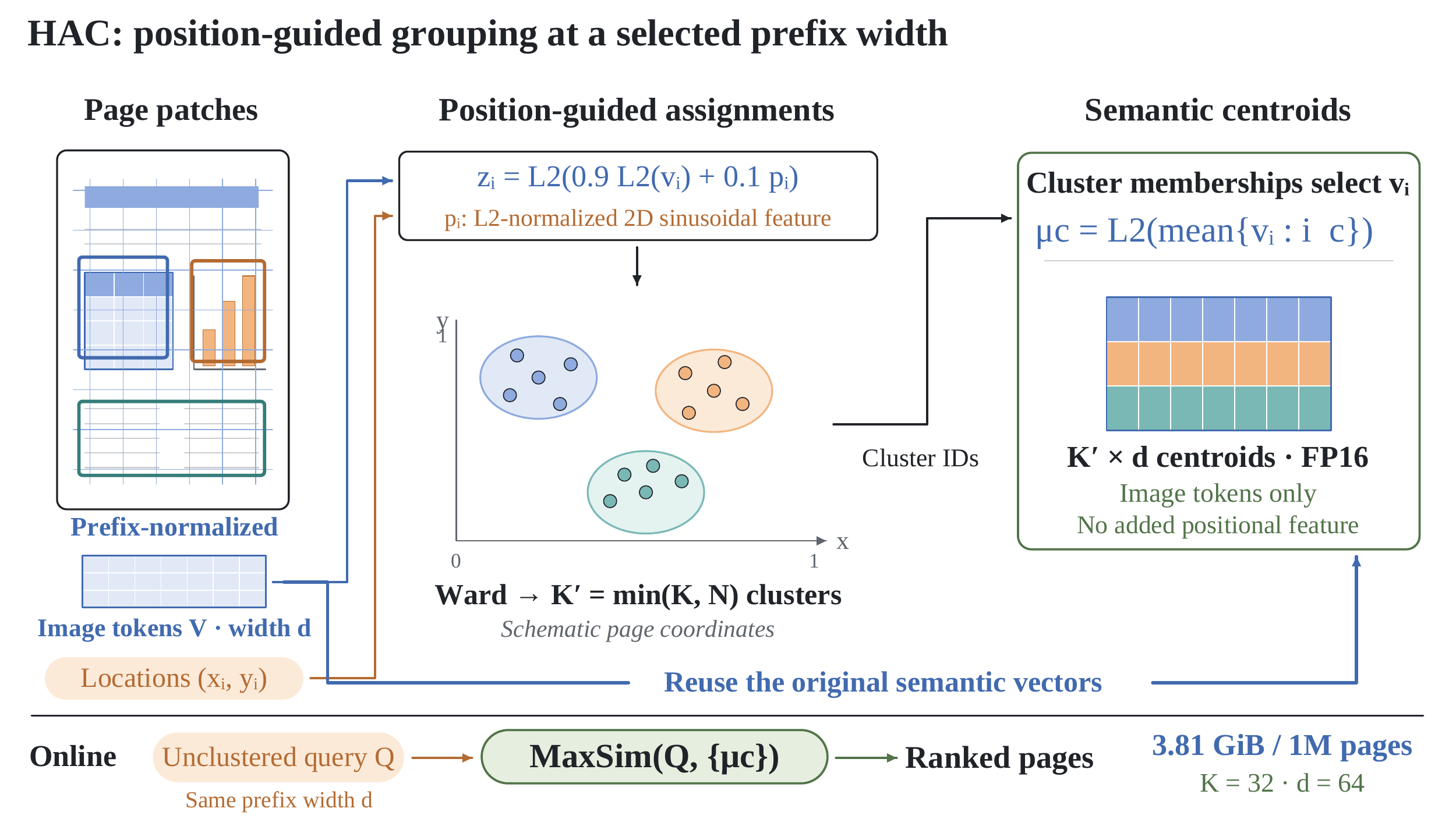}
\caption{\textbf{HAC separates position-guided assignment from semantic storage.} At a selected prefix width, normalized image vectors and normalized 2D sinusoidal features form Ward assignments. The blue bypass supplies original retrieval vectors for centroid averaging and normalization. Queries use the same width and score against the stored centroids in one MaxSim stage. The central scatter illustrates page coordinates, not measured clusters.}
\label{fig:hac_compression}
\end{figure}

Assignments are recomputed for each reported $(d,K)$ configuration. Prefix-MRL can derive narrower token representations from cached full-width projections without another backbone pass, but changing width still requires reclustering because both token geometry and the positional frequency grid depend on $d$. Algorithm~\ref{alg:hac} and Figure~\ref{fig:hac_compression} specify this indexing procedure.

\begin{algorithm}[htbp]
\caption{HAC Indexing at a Selected Prefix Width}
\label{alg:hac}
\footnotesize
\begin{algorithmic}[1]
\REQUIRE Projected page sequence $U$, image-token indices $\mathcal I$, merged image grid $(H,W)$
\REQUIRE Prefix width $d\in\mathcal D$, centroid budget $K$, positional weight $w_{\text{pos}}=0.1$
\ENSURE $K'=\min(K,|\mathcal I|)$ semantic centroids of width $d$
\STATE Select image tokens and normalize prefixes: $V\gets[\mathrm{L2}(U_{i,:d})]_{i\in\mathcal I}$
\STATE Set $N\gets|\mathcal I|$; verify $N=HW$; $K'\gets\min(K,N)$
\IF{$N\le K$}
    \STATE $\mathcal C\gets\{\{1\},\ldots,\{N\}\}$
\ELSE
    \STATE Construct raster coordinates and normalized $p_i$ by Eq.~(\ref{eq:pos})
    \STATE Form $Z=[z_i]_i$ by Eq.~(\ref{eq:zfeat}); $\mathcal C\gets\mathrm{Ward}(Z,K')$
\ENDIF
\STATE Compute each $\mu_c$ from $V$ by Eq.~(\ref{eq:centroid}) in FP32
\RETURN $\mathrm{FP16}([\mu_c]_{c\in\mathcal C})$
\end{algorithmic}
\end{algorithm}

\subsection{Index Storage Accounting}
Storage estimates count the embedding payload, excluding document IDs, offsets, serialization overhead, and model weights. The exported vectors use FP16, or two bytes per component; all storage estimates use $2^{30}$-byte units (GiB). For $P$ pages with $K$ vectors of width $d$,
\begin{equation}
S_{\mathrm{payload}}\;\text{(GiB)} = \frac{2PKd}{2^{30}}.
\end{equation}
Every page in the reported HAC runs has $N\ge K$ and therefore contributes exactly $K$ centroids. The primary configuration ($K=32,d=64$) uses 4 KiB per page, or 3.81 GiB per million pages: a $128\times$ reduction relative to EVIE's configured $1024\times256$ maximum visual-token payload in Table~\ref{tab:storage}. Width $d$ and centroid count $K$ are independently selectable indexing budgets whose costs multiply. Consequently, $(d=128,K=32)$ and $(d=64,K=64)$ both use 7.63 GiB per million pages; Section~\ref{sec:experiments} compares their retrieval accuracy.

\begin{table}[htbp]
\caption{Embedding payload per million pages at 16-bit precision. Uncompressed EVIE rows use the configured maximum of 1024 image vectors per page (\texttt{max\_num\_visual\_tokens}). Payload compression is relative to the first row's maximum visual-token payload. The primary HAC configuration is bold.}
\label{tab:storage}
\centering
\small\setlength{\tabcolsep}{2.5pt}
\begin{tabular}{lcccc}
\toprule
\textbf{Configuration} & \textbf{Vectors / Page} & \textbf{Dim ($d$)} & \textbf{GiB / 1M Pages} & \textbf{Compression} \\
\midrule
EVIE-4.5B Raw Max-Visual ($d=256$) & 1024 & 256 & 488.28 & $1.0\times$ \\
ColPali 1024-Vector Setting \citep{colpali2024} & 1024 & 128 & 244.14 & $2.0\times$ \\
EVIE-4.5B Raw Max-Visual ($d=128$) & 1024 & 128 & 244.14 & $2.0\times$ \\
EVIE-4.5B Raw Max-Visual ($d=64$) & 1024 & 64 & 122.07 & $4.0\times$ \\
\midrule
HAC ($d=128, K=32$) & 32 & 128 & 7.63 & $64.0\times$ \\
HAC ($d=64, K=64$) & 64 & 64 & 7.63 & $64.0\times$ \\
\textbf{HAC ($d=64, K=32$) [Ours]} & \textbf{32} & \textbf{64} & \textbf{3.81} & \textbf{128.0$\times$} \\
\bottomrule
\end{tabular}
\end{table}

\section{Experiments and Benchmark Results} \label{sec:experiments}

We evaluate retrieval quality across four benchmark suites and examine how embedding width and centroid count determine the accuracy--storage trade-off. The controlled comparisons below address the three contributions introduced in Sections~\ref{sec:governance}--\ref{sec:hac}: evidence-judged supervision, prefix-wise ranking transfer, and position-guided token compression.

\subsection{Benchmark Suite and Evaluation Protocol}

\paragraph{Benchmarks and aggregation.}
The evaluation comprises 138 tasks from four public suites. ViDoRe V3 \citep{vidore_v3_2025} contributes 48 domain--language tasks across eight public domains, with 14,514 queries and 115,512 language-expanded document evaluations over more than 19,000 unique page images. All V3 comparisons in this report use these eight public domains, averaging equally over six languages per domain and then over domains; private domains are excluded. ViDoRe V1 \citep{colpali2024} and V2 \citep{vidore_v2_2025} contain 10 tasks (3,943 queries) and four tasks (1,152 queries), respectively; JinaVDR \citep{jina_v4_2025} contributes 76 tasks and 43,375 queries. Table~\ref{tab:dataset_revisions} lists the individual ViDoRe dataset repositories and the revisions recorded in the local download metadata, including the MTEB-format V3 exports \citep{mmteb2025}. None of these four evaluation suites is included in the training data. The primary metric is nDCG@10, reported on a 0--100 scale. Each suite score is a task macro-average, and Avg4 is the unweighted mean of the four suite scores. This gives each suite equal weight despite their different task counts.

\paragraph{Evaluation protocol and baseline sources.}
The EVIE runs use the independently encoded, bidirectional representations of Section~\ref{sec:representation}, a maximum of 1024 image tokens, and single-stage MaxSim ranking without reranking. Raw runs retain image and context tokens; HAC runs retain image-token centroids only. External V3 scores in Tables~\ref{tab:vidore_v3} and~\ref{tab:suite_summary} use official model cards or benchmark result releases for the eight public domains; WeMM uses local measurements because corresponding official V3 results are unavailable. V1, V2, and JinaVDR scores, as well as all controlled comparisons, are evaluated locally. Avg4 is computed from these suite scores at the available source precision. Model-specific tokenization, resolution budgets, and training data may differ.

\paragraph{Relevance judgments and secondary metrics.}
ViDoRe V1 and JinaVDR use a single relevant page per query, whereas V2 and V3 include queries with multiple relevant pages. Recall measures coverage of those pages; MRR measures the rank of the first relevant result. These metrics answer different questions. Complete reported metrics and cutoffs are retained in Appendix~\ref{app:full_results}.

\paragraph{Storage accounting.}
All payloads in this section follow Section~\ref{sec:hac}: two bytes per vector component and GiB ($2^{30}$ bytes), excluding model weights, document IDs, offsets, and index-container overhead. HAC exports FP16 vectors. For raw EVIE, Table~\ref{tab:vidore_v3} uses the configured \texttt{max\_num\_visual\_tokens}=1024; for external multi-vector systems, it uses each model's published \texttt{max\_num\_visual\_tokens}.

\subsection{Main Retrieval Results Across Benchmarks}

\paragraph{Accuracy at explicit payload budgets.}
EVIE-8B reaches 66.75 nDCG@10 on the eight public ViDoRe V3 domains, 1.43 points above webAI-ColVec1.1-8b at 65.32, the strongest external baseline in Table~\ref{tab:vidore_v3}. EVIE-4.5B reaches 66.02 at $d=2048$ and 65.68 at $d=256$. The latter has a configured maximum visual-token payload of 488.28 GiB per million pages at $K_{\max}=1024$, compared with a published maximum visual-token payload of 2136.23 GiB for webAI-ColVec1.1-8b at $K_{\max}=1792$ and $d=640$. On this maximum visual-token basis, webAI's payload budget is $4.4\times$ larger, while EVIE-4.5B scores 0.36 points higher in nDCG@10.

At the compact end, HAC ($d=64,K=32$) reaches 59.58 at 3.81 GiB per million pages. It exceeds the WeMM-9B score of 57.42 by 2.16 points at half its vector payload. At the same 7.63 GiB payload as WeMM-9B, HAC ($d=64,K=64$) reaches 62.06, a gain of 4.64 points.

\begin{table}[htbp]
\caption{ViDoRe V3 nDCG@10 on eight public domains and embedding payload. $K$ is fixed for single-vector and HAC configurations and denotes the maximum visual-token count for raw multi-vector configurations. Payloads assume two-byte components and exclude index overhead.}
\label{tab:vidore_v3}
\centering
\scriptsize\setlength{\tabcolsep}{2.2pt}
\begin{tabular}{@{}p{0.37\linewidth}rrrrr@{}}
\toprule
Model & Params & Reported $K$ & $d$ & GiB / 1M & V3 \\
\midrule
\multicolumn{6}{@{}l}{\textit{Single-vector baselines (fixed $K$)}} \\
WeMM-2B \citep{wemm2026} & 2.1B & 1 & 2048 & 3.81 & 52.59 \\
Qwen3-VL-Embedding-2B \citep{qwen3vl_embed2025} & 2.1B & 1 & 2048 & 3.81 & 52.89 \\
WeMM-4B \citep{wemm2026} & 4.1B & 1 & 2560 & 4.77 & 54.70 \\
WeMM-9B \citep{wemm2026} & 9.0B & 1 & 4096 & 7.63 & 57.42 \\
\midrule
\multicolumn{6}{@{}l}{\textit{Multi-vector baselines (published $K_{\max}$)}} \\
VultronRetrieverFlash-0.8B \citep{vultron_flash_card} & 0.85B & 1792 & 320 & 1068.12 & 56.49 \\
tomoro-colqwen3-embed-4b \citep{tomoro_4b_card} & 4.0B & 1280 & 320 & 762.94 & 60.16 \\
nemotron-colembed-vl-4b-v2 \citep{nemotron_4b_card} & 4.8B & 2304 & 2560 & 10986.33 & 61.42 \\
colqwen3.5-4.5B-v3 \citep{colqwen35_v3_card} & 4.6B & 768 & 320 & 457.76 & 61.56 \\
tomoro-colqwen3-embed-8b \citep{tomoro_8b_card} & 8.0B & 1280 & 320 & 762.94 & 61.60 \\
nemotron-colembed-vl-8b-v2 \citep{nemotron_8b_card} & 8.8B & 2304 & 4096 & 17578.13 & 63.54 \\
VultronRetrieverCore-4.5B \citep{vultron_core_card} & 4.5B & 1792 & 320 & 1068.12 & 63.72 \\
webAI-ColVec1.1-4b \citep{webai_4b_card} & 4.5B & 1792 & 640 & 2136.23 & 64.24 \\
VultronRetrieverPrime-8B \citep{vultron_prime_card} & 8.4B & 1792 & 320 & 1068.12 & 64.72 \\
webAI-ColVec1.1-8b \citep{webai_8b_card} & 8.4B & 1792 & 640 & 2136.23 & 65.32 \\
\midrule
\multicolumn{6}{@{}l}{\textit{EVIE HAC measurements (fixed $K$)}} \\
HAC ($d=64,K=32$) & 4.5B & 32 & 64 & 3.81 & 59.58 \\
HAC ($d=64,K=64$) & 4.5B & 64 & 64 & 7.63 & 62.06 \\
HAC ($d=128,K=32$) & 4.5B & 32 & 128 & 7.63 & 61.40 \\
\midrule
\multicolumn{6}{@{}l}{\textit{EVIE raw scores (configured image-token $K_{\max}=1024$)}} \\
EVIE-Preview-4.5B (earlier release) & 4.5B & 1024 & 128 & 244.14 & 65.36 \\
EVIE-4.5B ($d=256$) & 4.5B & 1024 & 256 & 488.28 & 65.68 \\
EVIE-4.5B ($d=2048$) & 4.5B & 1024 & 2048 & 3906.25 & 66.02 \\
EVIE-8B (teacher) & 8.4B & 1024 & 4096 & 7812.50 & \textbf{66.75} \\
\bottomrule
\end{tabular}
\end{table}

\paragraph{Results across suites and domains.}
The WeMM-9B baseline obtains an Avg4 score of 71.51 (Table~\ref{tab:suite_summary}). The teacher and full-width student reach 79.51 and 78.98, improving Avg4 by 8.00 and 7.47 points, respectively, with gains on all four suites. HAC ($d=64,K=64$) reaches 75.57 at the same vector payload as WeMM-9B, while HAC ($d=64,K=32$) reaches 73.47 at half the payload, corresponding to Avg4 gains of 4.06 and 1.96 points. Both HAC configurations outperform WeMM-9B on V2, V3, and JinaVDR.

The single-vector and multi-vector configurations of jina-embeddings-v4 obtain Avg4 scores of 64.69 and 70.83, respectively, a difference of 6.14 points. EVIE-8B exceeds the multi-vector configuration by 8.68 Avg4 points, with higher scores on all four suites.

Relative to the full-width student, HAC ($d=64,K=32$) loses 3.26 points on V1, 7.08 on V2, 6.44 on V3, and 5.27 on JinaVDR, showing that the cost of compression varies across suites.

\begin{table}[htbp]
\caption{Main nDCG@10 results across four evaluation suites. V3 covers eight public domains and six languages per domain. Avg4 is the unweighted mean of the four suite scores.}
\label{tab:suite_summary}
\centering
\small\setlength{\tabcolsep}{3pt}
\begin{tabular}{@{}p{0.53\linewidth}rrrrr@{}}
\toprule
Model / configuration & V1 & V2 & V3 & JinaVDR & Avg4 \\
\midrule
WeMM-9B \citep{wemm2026} & 90.95 & 65.62 & 57.42 & 72.05 & 71.51 \\
nemotron-colembed-vl-8b-v2 \citep{nemotron_8b_card} & 92.27 & 66.74 & 63.54 & 81.41 & 75.99 \\
webAI-ColVec1.1-8b \citep{webai_8b_card} & 91.29 & 69.77 & 65.32 & 82.04 & 77.11 \\
jina-embeddings-v4 (single-vector) \citep{jina_v4_2025} & 84.67 & 55.74 & 49.61 & 68.74 & 64.69 \\
jina-embeddings-v4 (multi-vector) \citep{jina_v4_2025} & 89.73 & 60.43 & 56.99 & 76.17 & 70.83 \\
\midrule
EVIE-8B ($d=4096$) & 92.64 & 75.35 & 66.75 & 83.30 & \textbf{79.51} \\
EVIE-4.5B ($d=2048$) & 92.53 & 74.91 & 66.02 & 82.48 & 78.98 \\
EVIE-4.5B ($d=64$) & 92.16 & 73.18 & 64.51 & 81.00 & 77.71 \\
HAC ($d=64,K=64$) & 90.63 & 70.71 & 62.06 & 78.90 & 75.57 \\
HAC ($d=64,K=32$) & 89.27 & 67.83 & 59.58 & 77.21 & 73.47 \\
\bottomrule
\end{tabular}
\end{table}

Table~\ref{tab:v3_domains} resolves V3 into eight public domains, averaged over six languages per domain. The teacher--student gap ranges from 0.14 to 1.50 points. HAC scores 76.26 on Computer Science and 66.82 on Pharmaceuticals. Relative to the full-width student, the compression gap ranges from 1.89 points on Physics to 9.51 points on Human Resources, showing that sensitivity to the compact representation varies across domains.

\begin{table}[htbp]
\caption{ViDoRe V3 nDCG@10 on the eight public domains, macro-averaged over six languages per domain. HAC uses the final EVIE-4.5B checkpoint at $d=64$ and $K=32$.}
\label{tab:v3_domains}
\centering
\small\setlength{\tabcolsep}{6pt}
\begin{tabular}{@{}lrrr@{}}
\toprule
Domain & Teacher & Student ($d=2048$) & HAC ($d=64,K=32$) \\
\midrule
Computer Science & 81.86 & 81.72 & 76.26 \\
Energy & 72.51 & 72.32 & 67.11 \\
Finance (EN corpus) & 71.23 & 70.00 & 61.47 \\
Finance (FR corpus) & 56.40 & 54.90 & 46.21 \\
Human Resources & 69.29 & 67.82 & 58.31 \\
Industrial & 59.77 & 59.40 & 50.66 \\
Pharmaceuticals & 70.81 & 70.27 & 66.82 \\
Physics & 52.11 & 51.69 & 49.80 \\
\midrule
Overall & 66.75 & 66.02 & 59.58 \\
\bottomrule
\end{tabular}
\end{table}
\FloatBarrier

\subsection{Controlled Validation of the Three Contributions} \label{sec:controlled_validation}

\paragraph{Evidence-judged supervision.}
Table~\ref{tab:governance_ablation} evaluates three cumulative governance steps on the sampled-positive teacher branch before parameter averaging. Starting from the original supervision (G0), the sequence removes full-evidence candidates from negative slots (G1), admits them as positives (G2), and excludes partial-evidence candidates from the negative loss (G3). All four variants are trained independently from the same initialization checkpoint, with the same mined candidate pool, data-source mixture, query-group set and group order, optimizer and learning-rate schedule, effective batch size, number of optimizer updates, and random seed. Each scheduled query-group occurrence contributes one uniformly sampled positive. Positive expansion in G2 and G3 changes only the eligible positive-page set, leaving the group schedule and update budget unchanged. The only experimental factors varied are the three supervision switches
listed in the table.

\begin{table}[htbp]
\caption{Sequential ablation of evidence-judged supervision on one teacher branch. ``Full mask'' removes full-evidence candidates from negative slots; ``Expand'' adds them as positives; ``Partial mask'' excludes partial-evidence candidates. V3 and Avg4 report nDCG@10.}
\label{tab:governance_ablation}
\centering
\small\setlength{\tabcolsep}{5pt}
\begin{tabular}{@{}lcccrr@{}}
\toprule
Variant & Full mask & Expand & Partial mask & V3 & Avg4 \\
\midrule
G0: Original supervision & No & No & No & 62.36 &  75.57\\
G1: Remove full-evidence negatives & Yes & No & No & 64.23 & 77.71 \\
G2: Add full-evidence positives & Yes & Yes & No & 65.77 & 78.98 \\
G3: Exclude partial evidence & Yes & Yes & Yes & 66.33 & 79.17 \\
\bottomrule
\end{tabular}
\end{table}
Masking full-evidence negatives yields the largest incremental gain: G1 improves V3 by 1.87 points and Avg4 by 2.14 points over G0. Adding these pages as positives provides a further 1.54 and 1.27 points, respectively, showing additional benefit from explicit positive supervision after negative removal. Excluding partial evidence contributes another 0.56 points on V3 and 0.19 on Avg4. The complete sequence raises V3 from 62.36 to 66.33 and Avg4 from 75.57 to 79.17, for total gains of 3.97 and 3.60 points. Under this matched protocol, adjacent-row differences measure the incremental effect of each added governance step, conditional on the preceding steps.

\paragraph{Human agreement with evidence judgments.}
Table~\ref{tab:judge_audit} examines label quality on 900 candidates, with 300 from each judge tier (protocol in Appendix~\ref{app:extension_controls}). Three volunteer annotators independently apply the same full-, partial-, and no-evidence rubric, and their majority vote defines the human label for each candidate. Agreement with human labels is 92.67\% for full evidence, 80.67\% for partial evidence, and 95.67\% for no evidence. Direct confusion between full evidence and no evidence is 0.67\% in either direction, supporting the distinction between pages admitted as positives and pages retained as negatives. For partial-evidence pages, the G3 policy avoids treating incomplete answer support as definitive negative evidence. Because the audit samples equally from the three tiers, these rates describe tier-specific agreement rather than agreement under the original candidate distribution.

\begin{table}[htbp]
\caption{Stratified human audit of the evidence judge. Human labels are majority votes from three independent volunteer annotations. Rows group candidates by judge tier; each percentage is the corresponding majority-vote count divided by the 300 candidates sampled from that tier and is shown to two decimal places. Diagonal entries are tier-conditional agreement rates.}
\label{tab:judge_audit}
\centering
\small\setlength{\tabcolsep}{10pt}
\begin{tabular}{@{}lrrrr@{}}
\toprule
Judge tier & Audited $n$ & Human full (\%) & Partial (\%) & None (\%) \\
\midrule
Full evidence & 300 & 92.67 & 6.67 & 0.67 \\
Partial evidence & 300 & 9.67 & 80.67 & 9.67 \\
No evidence & 300 & 0.67 & 3.67 & 95.67 \\
\bottomrule
\end{tabular}
\end{table}

\paragraph{Prefix supervision and ranking transfer.}
Table~\ref{tab:distillation_ablation} compares post-hoc truncation of a maximum-width retrieval model (B0), retrieval supervision at all six prefixes (B1), query-to-page listwise distillation (B2), bidirectional listwise distillation (B3), and the complete student objective (B4). All five variants are controlled reruns with the same EVIE-Preview-4.5B initialization, training records and record order, optimizer and learning-rate schedule, number of update steps, effective batch size, and random seed. Thus, the only intended differences are the width-wise supervision and objective components stated in the table. At $d=64,128,2048$, adding query-to-page KL improves V3 by 0.65, 0.50, and 0.50 points over B1. Adding page-to-query KL yields a further 0.25 points at each reported width. The bidirectional variant therefore outperforms the query-to-page variant at all three widths. As defined in Eq.~(\ref{eq:kd}), bidirectionality concerns query-to-page and page-to-query normalization; both KL terms transfer distributions from teacher to student.

\begin{table}[htbp]
\caption{Matched student-training ablations, reporting V3 nDCG@10 at three representative widths. B0 applies retrieval supervision only at $d=2048$ and truncates the resulting representation at evaluation. B1--B3 successively introduce prefix retrieval supervision and the two KL directions, while B4 uses the complete objective in Eq.~(\ref{eq:student}), including margin transfer and the 128-dimensional anchor. B0--B4 use identical initialization, training records and order, optimizer schedule, update count, effective batch size, and seed; only the listed objective components differ.}
\label{tab:distillation_ablation}
\centering
\small\setlength{\tabcolsep}{5pt}
\begin{tabular}{@{}p{0.62\linewidth}rrr@{}}
\toprule
Training variant & $d=64$ & $d=128$ & $d=2048$ \\
\midrule
B0: Maximum-width retrieval; truncate at evaluation & 62.04 & 63.84 & 65.03 \\
B1: Retrieval supervision at all six prefixes & 62.54 & 63.90 & 65.06 \\
B2: B1 + query-to-page listwise KL & 63.19 & 64.40 & 65.56 \\
B3: B2 + page-to-query listwise KL & 63.44 & 64.65 & 65.81 \\
B4: Full objective & 64.51 & 65.27 & 66.02 \\
\bottomrule
\end{tabular}
\end{table}
Relative to B0 post-hoc truncation, direct prefix retrieval supervision in B1 improves V3 by 0.50 points at $d=64$, 0.06 at $d=128$, and 0.03 at $d=2048$. Direct prefix supervision is therefore most beneficial at the narrowest evaluated width. The full objective further improves over B3 by 1.07, 0.62, and 0.21 points at the three widths. Relative to prefix retrieval supervision alone, its total gains are 1.97 points at $d=64$, 1.37 at $d=128$, and 0.96 at $d=2048$. This pattern is consistent with the complete objective being most useful under tighter representation budgets. B4 evaluates margin transfer and anchor preservation jointly, so the table does not attribute their contributions separately. Table~\ref{tab:mrl_dims} complements these matched ablations by reporting all six widths of the final checkpoint.

\paragraph{Teacher construction.}
Table~\ref{tab:teacher_controls} reports the two teacher branches and their frozen parameter mixture. The sampled-positive and all-positive branches reach 66.33 and 66.23 on V3, with Avg4 scores of 79.17 and 79.06. Combining their parameters with weights of 0.4 and 0.6 raises V3 to 66.75 and Avg4 to 79.51, exceeding the stronger individual branch by 0.42 and 0.34 points, respectively. Parameter averaging therefore provides an additional improvement after evidence-judged supervision. The resulting frozen teacher supplies the ranking distributions and score margins used in student training.

\paragraph{HAC components on the final student.}
Table~\ref{tab:hac_components} holds the final student and $d=64$ fixed, first separating context-token removal from token pooling and then comparing pooling methods at $K=32$. Removing context tokens changes V3 from 64.51 to 64.39 and Avg4 from 77.71 to 77.63, losses of only 0.12 and 0.08 points. Compressing the image tokens with complete HAC then incurs losses of 4.81 and 4.16 points, respectively. Most of the raw-to-HAC accuracy loss at this operating point therefore arises from token pooling.

\begin{table}[htbp]
\caption{HAC component ablations on the final student at $d=64$, reporting nDCG@10. All pooling methods use image-token inputs and a 32-vector budget. $N$ denotes the number of image tokens and $N_{\rm all}$ the length of the valid raw page sequence. The Ward variants differ in the positional contribution to cluster assignment.}
\label{tab:hac_components}
\centering
\small\setlength{\tabcolsep}{5pt}
\begin{tabular}{@{}p{0.49\linewidth}ccrr@{}}
\toprule
Representation / pooling & Vectors & $w_{\rm pos}$ & V3 & Avg4 \\
\midrule
Raw, image and context tokens & $N_{\rm all}$ & --- & 64.51 & 77.71 \\
Raw, image tokens only & $N$ & --- & 64.39 & 77.63 \\
Grid pooling & 32 & --- & 49.37 & 63.87 \\
Semantic-only Ward & 32 & 0 & 58.12 & 72.59 \\
Position-guided Ward (HAC) & 32 & 0.1 & 59.58 & 73.47 \\
\bottomrule
\end{tabular}
\end{table}
At the same centroid budget, semantic-only Ward exceeds grid pooling by 8.75 points on V3 and 8.72 on Avg4, demonstrating the importance of semantic information in grouping page tokens. Introducing the positional term with $w_{\rm pos}=0.1$ adds a further 1.46 points on V3 and 0.88 on Avg4, reaching 59.58 and 73.47. Since the positional features guide cluster assignments while the stored centroids average the original semantic vectors, this gain is obtained without increasing embedding width or centroid count. The results support spatially guided semantic clustering at this operating point while retaining the single-stage MaxSim scoring interface described in Section~\ref{sec:hac}.
\FloatBarrier

\subsection{Accuracy--Storage Trade-offs and Budget Allocation}

\paragraph{Width selection from one student checkpoint.}
Table~\ref{tab:mrl_dims} reports six nested widths. Reducing $d$ from 2048 to 64 lowers V3 from 66.02 to 64.51 and Avg4 from 78.98 to 77.71: losses of 1.51 and 1.27 points, respectively, for a $32\times$ reduction in vector width. V3 and Avg4 increase monotonically across these measured widths; individual suites need not do so. At $d=128$, the student matches the EVIE-Preview checkpoint (65.27 vs.\ 65.36 on V3) while adding five further widths.

\begin{table}[htbp]
\caption{Prefix-MRL results, all nDCG@10, from one student checkpoint. Maximum visual-token payload uses EVIE's configured \texttt{max\_num\_visual\_tokens}=1024 at two bytes per component. Nested prefix extraction can reuse cached full-width projections; the representation property is distinct from the export history of each run.}
\label{tab:mrl_dims}
\centering
\small\setlength{\tabcolsep}{5pt}
\begin{tabular}{@{}rrrrrrr@{}}
\toprule
$d$ & GiB / 1M & V1 & V2 & V3 & JinaVDR & Avg4 \\
\midrule
64 & 122.07 & 92.16 & 73.18 & 64.51 & 81.00 & 77.71 \\
128 & 244.14 & 92.28 & 73.37 & 65.27 & 81.83 & 78.18 \\
256 & 488.28 & 92.20 & 73.83 & 65.68 & 82.09 & 78.45 \\
512 & 976.56 & 92.38 & 74.53 & 65.90 & 82.27 & 78.77 \\
1024 & 1953.13 & 92.39 & 74.53 & 65.94 & 82.43 & 78.82 \\
2048 & 3906.25 & 92.53 & 74.91 & 66.02 & 82.48 & 78.98 \\
\bottomrule
\end{tabular}
\end{table}

\paragraph{Where compression changes retrieval quality.}
Table~\ref{tab:hac_pareto} provides both the same-width raw configuration and the full-width student. On V3, narrowing the raw representation from $d=2048$ to $d=64$ lowers nDCG@10 from 66.02 to 64.51, a loss of 1.51 points, whereas compressing the $d=64$ tokens to $K=32$ HAC centroids lowers it further to 59.58, a loss of 4.93 points. Table~\ref{tab:hac_components} decomposes the latter into 0.12 points from context-token removal and 4.81 from image-token pooling.

\begin{table}[htbp]
\caption{Raw and HAC operating points, all nDCG@10. Raw payloads use EVIE's configured maximum of 1024 image tokens; HAC payloads use the fixed centroid count in the reported runs. Rows at the same $d$ help distinguish width changes from the subsequent compression procedure.}
\label{tab:hac_pareto}
\centering
\footnotesize\setlength{\tabcolsep}{4pt}
\begin{tabular}{@{}lrrrrrr@{}}
\toprule
Configuration & GiB / 1M & V1 & V2 & V3 & JinaVDR & Avg4 \\
\midrule
Raw ($d=2048$) & 3906.25 & 92.53 & 74.91 & 66.02 & 82.48 & 78.98 \\
Raw ($d=128$) & 244.14 & 92.28 & 73.37 & 65.27 & 81.83 & 78.18 \\
Raw ($d=64$) & 122.07 & 92.16 & 73.18 & 64.51 & 81.00 & 77.71 \\
\midrule
HAC ($d=128,K=32$) & 7.63 & 90.00 & 69.45 & 61.40 & 78.64 & 74.87 \\
HAC ($d=64,K=64$) & 7.63 & 90.63 & 70.71 & 62.06 & 78.90 & 75.57 \\
HAC ($d=64,K=32$) & 3.81 & 89.27 & 67.83 & 59.58 & 77.21 & 73.47 \\
\bottomrule
\end{tabular}
\end{table}

\begin{figure}[htbp]
\centering
\includegraphics[width=\linewidth]{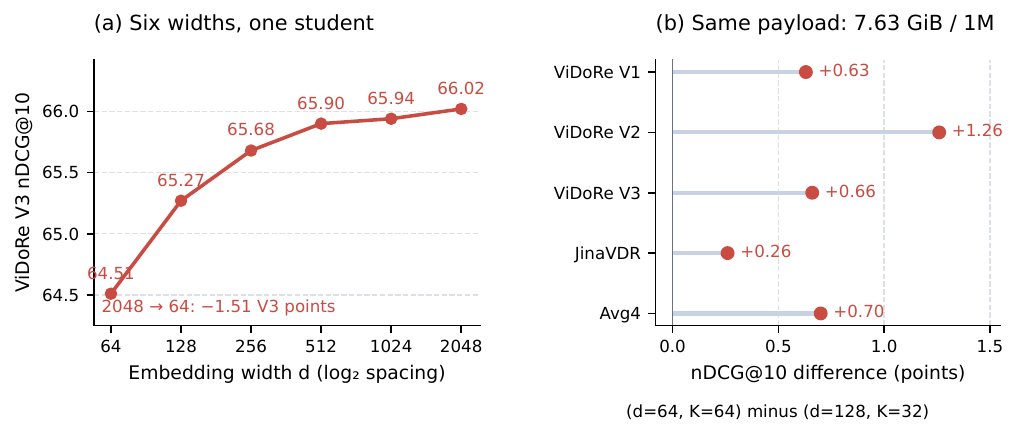}
\caption{\textbf{Measured width scaling and equal-payload differences.} (a) V3 nDCG@10 at six raw student widths; the endpoint loss is 1.51 V3 points. (b) Suite-level score differences between $(d=64,K=64)$ and $(d=128,K=32)$ at 7.63 GiB per million pages. Table~\ref{tab:budget_grid} reports the expanded comparison across three budgets.}
\label{fig:storage_levers}
\end{figure}

\paragraph{More vectors or wider vectors at equal storage.}
At 7.63 GiB per million pages, $(d=64,K=64)$ exceeds $(d=128,K=32)$ by 0.63, 1.26, 0.66, and 0.26 points on V1, V2, V3, and JinaVDR, respectively (Figure~\ref{fig:storage_levers}), with an Avg4 gain of 0.70 points. Table~\ref{tab:budget_grid} extends this comparison to three payload budgets, varying $d\in\{64,128,256\}$ and $K$ while holding their product fixed within each budget.

\begin{table}[htbp]
\caption{Expanded equal-payload comparison for HAC on the final EVIE-4.5B checkpoint. Each budget fixes $Kd$ while varying embedding width and centroid count. Payloads assume $K$ centroids per page at two bytes per component; V3 and Avg4 report nDCG@10.}
\label{tab:budget_grid}
\centering
\small\setlength{\tabcolsep}{10pt}
\begin{tabular}{@{}rrrrr@{}}
\toprule
GiB / 1M & $d$ & $K$ & V3 & Avg4 \\
\midrule
3.81 & 64 & 32 & 59.58 & 73.47 \\
3.81 & 128 & 16 & 58.70 & 72.38 \\
3.81 & 256 & 8 & 55.19 & 67.63 \\
\midrule
7.63 & 64 & 64 & 62.06 & 75.57 \\
7.63 & 128 & 32 & 61.40 & 74.87 \\
7.63 & 256 & 16 & 59.28 & 72.63 \\
\midrule
15.26 & 64 & 128 & 63.40 & 76.86 \\
15.26 & 128 & 64 & 63.33 & 76.56 \\
15.26 & 256 & 32 & 61.98 & 75.33 \\
\bottomrule
\end{tabular}
\end{table}
At all three budgets, allocating the payload to more 64-dimensional centroids yields the highest V3 and Avg4 scores among the tested configurations. Relative to $d=128$, the $d=64$ configurations improve V3 by 0.88, 0.66, and 0.07 points at 3.81, 7.63, and 15.26 GiB, respectively; the corresponding Avg4 gains are 1.09, 0.70, and 0.30 points. They also outperform the $d=256$ configurations, with V3 gains ranging from 1.42 to 4.39 points. The advantage over wider vectors decreases as the payload budget increases. These results favor retaining more centroids at a narrower width for the tested EVIE-4.5B budgets.

\paragraph{Accuracy--storage overview.}
Figure~\ref{fig:pareto_curve} gives a descriptive overview of accuracy against the heterogeneous payload quantities reported for EVIE and external systems. Tables~\ref{tab:mrl_dims} and~\ref{tab:hac_pareto} provide the detailed operating points of the final student, including intermediate widths omitted from the overview. These settings support selecting a width and centroid budget from one student checkpoint; the Preview and teacher points in Figure~\ref{fig:pareto_curve} refer to separate checkpoints.

\FloatBarrier

\subsection{System Measurements and Evidence-Level Analysis}

\paragraph{Serving latency and throughput.}
Table~\ref{tab:system_measurements} reports EVIE-4.5B and WeMM-9B on a 1000-page stratified ViDoRe V3 corpus (seed 42, 125 pages per domain) with all 14{,}514 queries. Latency is batch 1; QPS is batch 16; Search scores already-encoded query vectors. The four EVIE rows share one Prefix-MRL $d{=}64$ encoding. Relative to raw $d{=}64$, HAC ($K{=}32$) reduces median search time from 0.909 to 0.160 ms ($5.7\times$) and GPU-resident index memory by $32\times$; its batch-16 throughput of 115.73 QPS is 20\% higher than that of WeMM-9B. Full-suite accuracy is in Tables~\ref{tab:mrl_dims} and~\ref{tab:hac_pareto}.

\begin{table}[H]
\caption{1000 stratified ViDoRe V3 pages, 14{,}514 queries, one NVIDIA H20, MVT${=}1024$. Enc.\ and Search are medians; Search scores already-encoded query vectors. E2E is batch 1; QPS is end-to-end at batch 16. Cascade: HAC $K{=}32$ top-100 then raw $d{=}64$ MaxSim. Mem.: GPU-resident index (GiB).}
\label{tab:system_measurements}
\centering
\footnotesize\setlength{\tabcolsep}{3.5pt}
\begin{tabular}{@{}lrrrrrr@{}}
\toprule
System & Enc.\ (ms) & Search (ms) & E2E p50 & E2E p95 & QPS & Mem. \\
\midrule
WeMM-9B & 106.78 & \textbf{0.125} & 106.90 & 110.29 & 96.45 & \textbf{0.0076} \\
EVIE raw ($d{=}64$) & \textbf{99.67} & 0.909 & 100.60 & 104.35 & 105.26 & 0.2444 \\
EVIE HAC ($d{=}64,K{=}32$) & \textbf{99.67} & 0.160 & 99.83 & 103.62 & \textbf{115.73} & \textbf{0.0076} \\
EVIE HAC ($d{=}64,K{=}64$) & \textbf{99.67} & 0.150 & \textbf{99.82} & \textbf{103.60} & 115.65 & 0.0153 \\
EVIE cascade ($K{=}32$ top-100) & \textbf{99.67} & 0.335 & 100.01 & 103.79 & 113.27 & 0.2520 \\
\bottomrule
\end{tabular}
\end{table}

\begin{figure}[H]
\centering
\includegraphics[width=\linewidth]{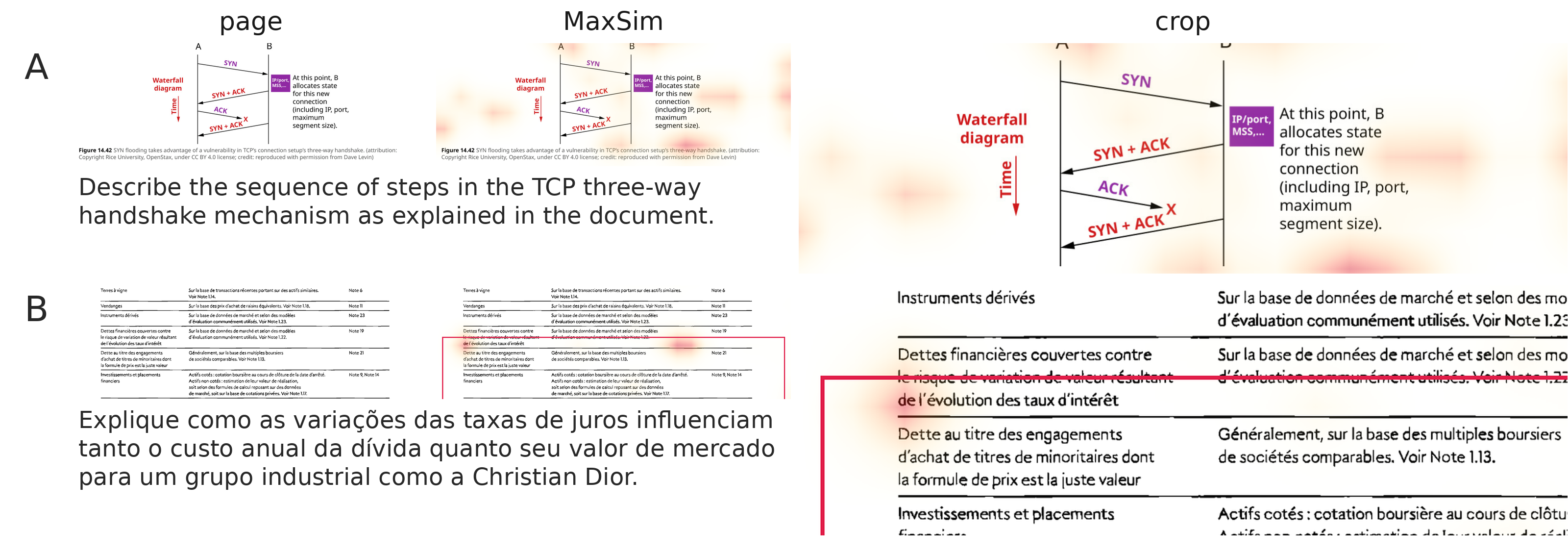}
\caption{\textbf{Retained and lost retrieval evidence under HAC.} For cases A and B: page, MaxSim map, and crop of the top token's HAC cluster.}
\label{fig:evidence_cases}
\end{figure}

\paragraph{Evidence cases under HAC.}
On the same index, 3{,}588 queries have a relevant page: 2{,}521 stay in the top three under both raw $d{=}64$ and HAC $K{=}32$, and 49 have raw rank $\le 3$ with HAC rank $>10$. Figure~\ref{fig:evidence_cases} shows one of each. A ranks 1 in both indexes; B is raw 3 and HAC 37. Yellow--red is the upper MaxSim mass; the box is the HAC cluster of the top token.

\FloatBarrier

\section{Conclusion and Future Work}
\label{sec:conclusion}
\label{sec:future_work}

In this report, we introduced EVIE, a family of native visual document retrieval models combining evidence-judged supervision, bidirectional teacher--student learning with Prefix-MRL, and hierarchical agglomerative index compression (HAC). The framework connects page-level relevance learning with flexible embedding widths and compact semantic centroids for single-stage MaxSim retrieval. Evaluations across 138 tasks show a favorable accuracy--storage trade-off, with EVIE-8B reaching 66.75 nDCG@10 on ViDoRe V3 and EVIE-4.5B with HAC retaining 59.58 at 3.81 GiB of vector payload per million pages. These results support adapting fine-grained visual document retrieval to different storage budgets. In the future, we will extend EVIE to image queries, including screenshots and document crops, alongside text queries. We aim to align both query modalities with page representations in a shared retrieval space, allowing them to reuse the same compressed document index and reducing duplicate vector storage.

\subsection*{Ethics Statement}
This work uses existing public document-retrieval datasets. The recorded training-data preprocessing screens out empty queries and images; the evidence judgments in Section~\ref{sec:governance} assess relevance and determine positive expansion and negative masking. These validity and relevance checks do not constitute a comprehensive privacy or harmful-content audit. Public document pages may contain names or other personally identifiable information, and we do not claim that such content is entirely absent. Use and redistribution remain subject to each dataset's license and any underlying document restrictions; public availability and the model license do not override those terms. Corpus-specific privacy review and access controls remain necessary for deployment.

{\small
\bibliography{iclr2027_conference}
\bibliographystyle{iclr2027_conference}
}

\clearpage
\appendix
\section{Detailed Multi-Metric Results} \label{app:full_results}

Tables~\ref{tab:full_138}--\ref{tab:full_138_hac} report nDCG, Recall, MAP, and MRR at cutoffs 1, 5, and 10 for three representative EVIE configurations. The V3 columns use the same 48 public domain--language tasks as Section~\ref{sec:experiments}. For ViDoRe V1 and JinaVDR, each query has one binary relevant page. For ViDoRe V2 and V3, qrel entries with gain $g>0$ are relevant for Recall, MAP, and MRR, while nDCG retains the graded gains using $2^g-1$ with logarithmic rank discount. At cutoff $k$, MAP is normalized by $\min(k,|R_q|)$. Metrics are averaged over queries within each task and macro-averaged over tasks within each suite; Avg4 is the unweighted mean of the four suite scores. All scores are reported on a 0--100 scale. MAP and MRR coincide at $k=1$ for all suites and at every cutoff for single-relevant-page tasks.

\begin{table}[htbp]
\caption{Complete reported retrieval metrics for the EVIE-8B teacher.}
\label{tab:full_138}
\centering
\small\setlength{\tabcolsep}{7pt}
\begin{tabular}{@{}lrrrrr@{}}
\toprule
Metric & ViDoRe V1 & ViDoRe V2 & ViDoRe V3 & JinaVDR & Avg4 \\
\midrule
nDCG@1 & 88.51 & 75.58 & 62.14 & 75.50 & 75.43 \\
nDCG@5 & 92.18 & 74.23 & 64.16 & 82.45 & 78.26 \\
nDCG@10 & 92.64 & 75.35 & 66.75 & 83.30 & 79.51 \\
\midrule
Recall@1 & 88.51 & 37.87 & 30.79 & 75.50 & 58.17 \\
Recall@5 & 95.07 & 66.65 & 58.69 & 88.07 & 77.12 \\
Recall@10 & 96.43 & 76.34 & 70.10 & 90.65 & 83.38 \\
\midrule
MAP@1 & 88.51 & 75.99 & 65.94 & 75.50 & 76.49 \\
MAP@5 & 91.21 & 67.30 & 55.82 & 80.56 & 73.72 \\
MAP@10 & 91.40 & 66.81 & 56.32 & 80.92 & 73.86 \\
\midrule
MRR@1 & 88.51 & 75.99 & 65.94 & 75.50 & 76.49 \\
MRR@5 & 91.21 & 82.85 & 75.39 & 80.56 & 82.50 \\
MRR@10 & 91.40 & 83.23 & 76.09 & 80.92 & 82.91 \\
\bottomrule
\end{tabular}
\end{table}

\begin{table}[htbp]
\caption{Complete reported retrieval metrics for the EVIE-4.5B student at $d=2048$.}
\label{tab:full_138_student}
\centering
\small\setlength{\tabcolsep}{7pt}
\begin{tabular}{@{}lrrrrr@{}}
\toprule
Metric & ViDoRe V1 & ViDoRe V2 & ViDoRe V3 & JinaVDR & Avg4 \\
\midrule
nDCG@1 & 88.35 & 72.76 & 61.16 & 74.26 & 74.13 \\
nDCG@5 & 92.07 & 73.38 & 63.39 & 81.57 & 77.60 \\
nDCG@10 & 92.53 & 74.91 & 66.02 & 82.48 & 78.98 \\
\midrule
Recall@1 & 88.35 & 36.73 & 30.14 & 74.26 & 57.37 \\
Recall@5 & 94.95 & 65.59 & 58.06 & 87.39 & 76.50 \\
Recall@10 & 96.36 & 76.56 & 69.67 & 90.17 & 83.19 \\
\midrule
MAP@1 & 88.35 & 73.18 & 65.03 & 74.26 & 75.20 \\
MAP@5 & 91.09 & 66.52 & 55.02 & 79.61 & 73.06 \\
MAP@10 & 91.29 & 66.25 & 55.50 & 79.99 & 73.25 \\
\midrule
MRR@1 & 88.35 & 73.18 & 65.03 & 74.26 & 75.20 \\
MRR@5 & 91.09 & 81.66 & 74.76 & 79.61 & 81.78 \\
MRR@10 & 91.29 & 82.11 & 75.43 & 79.99 & 82.20 \\
\bottomrule
\end{tabular}
\end{table}

\begin{table}[htbp]
\caption{Complete reported retrieval metrics for the HAC at $d=64,K=32$.}
\label{tab:full_138_hac}
\centering
\small\setlength{\tabcolsep}{7pt}
\begin{tabular}{@{}lrrrrr@{}}
\toprule
Metric & ViDoRe V1 & ViDoRe V2 & ViDoRe V3 & JinaVDR & Avg4 \\
\midrule
nDCG@1 & 82.95 & 65.82 & 54.29 & 67.50 & 67.64 \\
nDCG@5 & 88.74 & 66.19 & 56.98 & 75.89 & 71.95 \\
nDCG@10 & 89.27 & 67.83 & 59.58 & 77.21 & 73.47 \\
\midrule
Recall@1 & 82.95 & 31.16 & 26.35 & 67.50 & 51.99 \\
Recall@5 & 93.30 & 60.17 & 52.82 & 82.84 & 72.28 \\
Recall@10 & 94.94 & 72.01 & 64.00 & 86.90 & 79.46 \\
\midrule
MAP@1 & 82.95 & 65.82 & 58.26 & 67.50 & 68.63 \\
MAP@5 & 87.20 & 58.45 & 48.34 & 73.56 & 66.89 \\
MAP@10 & 87.42 & 57.65 & 48.60 & 74.11 & 66.95 \\
\midrule
MRR@1 & 82.95 & 65.82 & 58.26 & 67.50 & 68.63 \\
MRR@5 & 87.20 & 75.25 & 68.89 & 73.56 & 76.22 \\
MRR@10 & 87.42 & 76.12 & 69.73 & 74.11 & 76.84 \\
\bottomrule
\end{tabular}
\end{table}

\FloatBarrier

\section{Dataset Provenance} \label{app:source_provenance}

\paragraph{Dataset snapshots.}
Table~\ref{tab:dataset_revisions} identifies the 22 ViDoRe repositories used by the local evaluator: ten V1 datasets, four V2 datasets, and eight public V3 domain datasets. The V3 exports provide six query languages per domain; the evaluator shares the English corpus across those language tasks. All listed configurations use the \texttt{test} split. Revisions were recovered from the per-file Hugging Face download metadata accompanying the local snapshots, rather than from the repositories' current default branches. The accompanying \texttt{dataset\_revisions.json} retains full commit hashes and file-level download identifiers.

\begin{table}[htbp]
\caption{ViDoRe dataset sources and archived revisions. Every repository is under the \texttt{vidore/} namespace. Abbreviated commit identifiers link to the full immutable revision. V1 uses the default data configuration; V2 uses \texttt{corpus}, \texttt{queries}, and \texttt{qrels}; V3 uses \texttt{english-corpus} and the six language-specific query/qrels configurations.}
\label{tab:dataset_revisions}
\centering
\scriptsize\setlength{\tabcolsep}{4pt}
\begin{tabular}{@{}lp{0.66\linewidth}l@{}}
\toprule
Suite & Dataset repository & Revision \\
\midrule
V1 & \href{https://huggingface.co/datasets/vidore/arxivqa_test_subsampled}{\nolinkurl{arxivqa_test_subsampled}} & \href{https://huggingface.co/datasets/vidore/arxivqa_test_subsampled/tree/b8a106812c8682bab08935cf5d1b4566c82562de}{\texttt{b8a106812c86}} \\
V1 & \href{https://huggingface.co/datasets/vidore/docvqa_test_subsampled}{\nolinkurl{docvqa_test_subsampled}} & \href{https://huggingface.co/datasets/vidore/docvqa_test_subsampled/tree/49bf8f13e13c41dd8cdb0cae5314e31c1da1e0d6}{\texttt{49bf8f13e13c}} \\
V1 & \href{https://huggingface.co/datasets/vidore/infovqa_test_subsampled}{\nolinkurl{infovqa_test_subsampled}} & \href{https://huggingface.co/datasets/vidore/infovqa_test_subsampled/tree/f793e830aaeae1ceb8a2df626fc555b3fd04d3db}{\texttt{f793e830aaea}} \\
V1 & \href{https://huggingface.co/datasets/vidore/shiftproject_test}{\nolinkurl{shiftproject_test}} & \href{https://huggingface.co/datasets/vidore/shiftproject_test/tree/6e6223b3839a3f4e62d676e70a1b715ee520cc66}{\texttt{6e6223b3839a}} \\
V1 & \href{https://huggingface.co/datasets/vidore/syntheticDocQA_artificial_intelligence_test}{\nolinkurl{syntheticDocQA_artificial_intelligence_test}} & \href{https://huggingface.co/datasets/vidore/syntheticDocQA_artificial_intelligence_test/tree/5694fec64c57f7380f918435f9234b96d817b82a}{\texttt{5694fec64c57}} \\
V1 & \href{https://huggingface.co/datasets/vidore/syntheticDocQA_energy_test}{\nolinkurl{syntheticDocQA_energy_test}} & \href{https://huggingface.co/datasets/vidore/syntheticDocQA_energy_test/tree/a2f2f358463b03b84506849460ab5094a358526c}{\texttt{a2f2f358463b}} \\
V1 & \href{https://huggingface.co/datasets/vidore/syntheticDocQA_government_reports_test}{\nolinkurl{syntheticDocQA_government_reports_test}} & \href{https://huggingface.co/datasets/vidore/syntheticDocQA_government_reports_test/tree/91cf66572d89c9cccac0661de227acaf04b44f64}{\texttt{91cf66572d89}} \\
V1 & \href{https://huggingface.co/datasets/vidore/syntheticDocQA_healthcare_industry_test}{\nolinkurl{syntheticDocQA_healthcare_industry_test}} & \href{https://huggingface.co/datasets/vidore/syntheticDocQA_healthcare_industry_test/tree/d973f3da3e60c8eaf566efa2f7d2a1515cba40ed}{\texttt{d973f3da3e60}} \\
V1 & \href{https://huggingface.co/datasets/vidore/tabfquad_test_subsampled}{\nolinkurl{tabfquad_test_subsampled}} & \href{https://huggingface.co/datasets/vidore/tabfquad_test_subsampled/tree/16c8e633612fbda7400bfcbbc31d61a7534f580f}{\texttt{16c8e633612f}} \\
V1 & \href{https://huggingface.co/datasets/vidore/tatdqa_test}{\nolinkurl{tatdqa_test}} & \href{https://huggingface.co/datasets/vidore/tatdqa_test/tree/b46fea43695e14697510104a3331d9e88683a416}{\texttt{b46fea43695e}} \\
\midrule
V2 & \href{https://huggingface.co/datasets/vidore/biomedical_lectures_v2}{\nolinkurl{biomedical_lectures_v2}} & \href{https://huggingface.co/datasets/vidore/biomedical_lectures_v2/tree/c4754665734e38742b191f0c28d504e8558d0462}{\texttt{c4754665734e}} \\
V2 & \href{https://huggingface.co/datasets/vidore/economics_reports_v2}{\nolinkurl{economics_reports_v2}} & \href{https://huggingface.co/datasets/vidore/economics_reports_v2/tree/76fe40166ba07b1bf50457f5c6057cacdd045f10}{\texttt{76fe40166ba0}} \\
V2 & \href{https://huggingface.co/datasets/vidore/esg_reports_v2}{\nolinkurl{esg_reports_v2}} & \href{https://huggingface.co/datasets/vidore/esg_reports_v2/tree/87538b12b20b67a2b4326638921301f87f0cbaf0}{\texttt{87538b12b20b}} \\
V2 & \href{https://huggingface.co/datasets/vidore/esg_reports_human_labeled_v2}{\nolinkurl{esg_reports_human_labeled_v2}} & \href{https://huggingface.co/datasets/vidore/esg_reports_human_labeled_v2/tree/5a338c329bf1608ac46ac2808060d44bcd92d521}{\texttt{5a338c329bf1}} \\
\midrule
V3 & \href{https://huggingface.co/datasets/vidore/vidore_v3_computer_science_mteb_format}{\nolinkurl{vidore_v3_computer_science_mteb_format}} & \href{https://huggingface.co/datasets/vidore/vidore_v3_computer_science_mteb_format/tree/fb7fb69f81f7db62790f40494124b8ad22b424ab}{\texttt{fb7fb69f81f7}} \\
V3 & \href{https://huggingface.co/datasets/vidore/vidore_v3_energy_mteb_format}{\nolinkurl{vidore_v3_energy_mteb_format}} & \href{https://huggingface.co/datasets/vidore/vidore_v3_energy_mteb_format/tree/84fca99e5978604bae30f2436eacb6dbaa0532e9}{\texttt{84fca99e5978}} \\
V3 & \href{https://huggingface.co/datasets/vidore/vidore_v3_finance_en_mteb_format}{\nolinkurl{vidore_v3_finance_en_mteb_format}} & \href{https://huggingface.co/datasets/vidore/vidore_v3_finance_en_mteb_format/tree/fa78cb14152b3dde8c5defdc4e3ddf50de69dfeb}{\texttt{fa78cb14152b}} \\
V3 & \href{https://huggingface.co/datasets/vidore/vidore_v3_finance_fr_mteb_format}{\nolinkurl{vidore_v3_finance_fr_mteb_format}} & \href{https://huggingface.co/datasets/vidore/vidore_v3_finance_fr_mteb_format/tree/8a2adfda85a7967c7252129703d9b3c7c9f038a9}{\texttt{8a2adfda85a7}} \\
V3 & \href{https://huggingface.co/datasets/vidore/vidore_v3_hr_mteb_format}{\nolinkurl{vidore_v3_hr_mteb_format}} & \href{https://huggingface.co/datasets/vidore/vidore_v3_hr_mteb_format/tree/bc7d43d64815ed30f664168c8052106484aba7fd}{\texttt{bc7d43d64815}} \\
V3 & \href{https://huggingface.co/datasets/vidore/vidore_v3_industrial_mteb_format}{\nolinkurl{vidore_v3_industrial_mteb_format}} & \href{https://huggingface.co/datasets/vidore/vidore_v3_industrial_mteb_format/tree/f732b725cf4a70803210edfe265a04f8bd5328f6}{\texttt{f732b725cf4a}} \\
V3 & \href{https://huggingface.co/datasets/vidore/vidore_v3_pharmaceuticals_mteb_format}{\nolinkurl{vidore_v3_pharmaceuticals_mteb_format}} & \href{https://huggingface.co/datasets/vidore/vidore_v3_pharmaceuticals_mteb_format/tree/237ed4f43c7fb3c4df07ec4e9dd0a4366be555b0}{\texttt{237ed4f43c7f}} \\
V3 & \href{https://huggingface.co/datasets/vidore/vidore_v3_physics_mteb_format}{\nolinkurl{vidore_v3_physics_mteb_format}} & \href{https://huggingface.co/datasets/vidore/vidore_v3_physics_mteb_format/tree/2c18ef90ab3ef93a9d86ecc6521cdae2a29f8300}{\texttt{2c18ef90ab3e}} \\
\bottomrule
\end{tabular}
\end{table}

\FloatBarrier

\section{Additional Controlled Comparisons} \label{app:extension_controls}

V3 scores follow the protocol for eight public domains in Section~\ref{sec:experiments}.

\paragraph{Human audit of evidence judgments.}
The audit contains 900 candidates sampled equally from the three judge tiers. Three volunteer annotators independently assign every candidate to one of the same three answer-completeness categories used by the judge. A page receives a full-evidence label when it independently supports a complete answer, a partial-evidence label when it contains useful but insufficient answer support, and a no-evidence label when it does not support the answer. The majority vote of the three annotations defines the human label used in Table~\ref{tab:judge_audit}. Each table cell is the corresponding majority-vote count divided by the 300 candidates sampled from that judge tier.

\paragraph{Positional-weight provenance.}
The weight $w_{\mathrm{pos}}=0.1$ was fixed from a sweep over $\{0,0.1,\ldots,0.9\}$ using an earlier student checkpoint at $d=128,K=32$, then applied unchanged to the final student. Table~\ref{tab:hac_components} evaluates this fixed setting without retuning it for the final checkpoint.

\paragraph{Teacher construction.}
The final teacher combines the sampled-positive and all-positive branches through parameter averaging, $\theta_T=0.4\theta_{\mathrm{sampled}}+0.6\theta_{\mathrm{all}}$. The merged checkpoint is frozen for student supervision; Table~\ref{tab:teacher_controls} reports the branch and mixture results.
At the evaluated $0.4/0.6$ weighting, the mixture exceeds the stronger branch by 0.42 V3 points and 0.34 Avg4 points. This single mixture comparison does not establish the optimum weights or an improvement averaged over training seeds.
\begin{table}[htbp]
\caption{Teacher construction at $d=4096$ on the same 138 tasks. The final mixture weights the sampled-positive and all-positive branches by 0.4 and 0.6, respectively.}
\label{tab:teacher_controls}
\centering
\small\setlength{\tabcolsep}{8pt}
\begin{tabular}{@{}lrr@{}}
\toprule
Teacher checkpoint & V3 & Avg4 \\
\midrule
Sampled-positive branch & 66.33 & 79.17 \\
All-positive branch & 66.23 & 79.06 \\
Final mixture ($0.4/0.6$) & \textbf{66.75} & \textbf{79.51} \\
\bottomrule
\end{tabular}
\end{table}

\paragraph{Compression transfer across encoders.}
Table~\ref{tab:hac_transfer} compares semantic-only Ward and position-guided HAC on EVIE-4.5B and two external multi-vector encoders. Spatial coordinates follow each encoder's native processor grid.

\begin{table}[htbp]
\caption{Compression transfer on V3 (nDCG@10). Ward and HAC use the same image tokens, width, $K=32$, and precision within each encoder and store semantic centroids. Widths differ across encoders, so payloads are not matched.}
\label{tab:hac_transfer}
\centering
\small\setlength{\tabcolsep}{5pt}
\begin{tabular}{@{}lrrrr@{}}
\toprule
Encoder & $d$ & Raw image-only & Ward ($w=0$) & HAC ($w=0.1$) \\
\midrule
EVIE-4.5B (final) & 64 & 64.39 & 58.12 & 59.58 \\
Jina v4 (multi-vector) & 128 & 56.38 & 49.60 & 50.88 \\
Nemotron-ColEmbed-V2-4B & 2560 & 61.62 & 57.98 & 58.76 \\
\bottomrule
\end{tabular}
\end{table}
HAC improves over Ward by 1.46, 1.28, and 0.78 points on EVIE, Jina, and Nemotron, respectively, with raw-to-HAC losses of 4.81, 5.50, and 2.86 points. The positional gain extends to both external encoders at this budget.

\paragraph{Governance and compression sensitivity.}
Table~\ref{tab:governance_compression} compares original and governed student supervision at $d=64$. Compression sensitivity is measured by the within-model score loss from raw image-only retrieval to HAC.

\begin{table}[htbp]
\caption{Governance and compression on V3 (nDCG@10) at $d=64$. Both modes use image tokens only; HAC uses $K=32$ and $w=0.1$. Loss is raw minus HAC.}
\label{tab:governance_compression}
\centering
\small\setlength{\tabcolsep}{8pt}
\begin{tabular}{@{}lrrr@{}}
\toprule
Governance treatment & Raw image-only V3 & HAC V3 ($K=32$) & Loss \\
\midrule
Original supervision & 61.38 & 54.80 & 6.58 \\
Full governance & 64.39 & 59.58 & 4.81 \\
\bottomrule
\end{tabular}
\end{table}
Full governance raises raw and HAC scores by 3.01 and 4.78 points, respectively. Compression loss falls from 6.58 to 4.81 points, giving an interaction estimate of $6.58-4.81=1.77$ points. The larger gain after compression is consistent with reduced sensitivity at this budget; without uncertainty intervals, the interaction remains descriptive.
\FloatBarrier

\end{document}